\documentclass[runningheads]{llncs}
\usepackage[T1]{fontenc}
\usepackage{setspace}              
\usepackage{bm}
\usepackage{amssymb}             
\usepackage{tikz}            
\usepackage{quantikz} 
\usepackage{enumitem}
\usepackage{graphicx} 
\usepackage{algpseudocode}
\usepackage{algorithm}
\usepackage{algorithmicx}

\usepackage{graphicx}

\usepackage[utf8]{inputenc}
\usepackage[english]{babel}
\usepackage{amssymb}
\usepackage{physics}

\usepackage{hyperref}
\usepackage[table]{xcolor}
\usepackage{colortbl}
\usepackage{booktabs}
\usepackage{tabularx}
\usepackage{caption}
\usepackage{subcaption}

\usepackage{selinput}
\SelectInputMappings{
  adieresis={ä}, 
  acute={´},
}

\usepackage[english]{babel}
\usepackage[T1]{fontenc}
\usepackage{lipsum}
\usepackage{mdframed}
\usepackage{booktabs}
\usepackage{adjustbox}
\usepackage{longtable}
 
\usepackage{placeins}
\newcolumntype{L}{>{\raggedright\arraybackslash}X}
\newcolumntype{C}{>{\centering\arraybackslash}X}
\newcolumntype{R}{>{\raggedleft\arraybackslash}X}

\usepackage{xcolor}
\usepackage[table]{xcolor}
\usepackage{booktabs}
\usepackage{listings}
\usepackage{siunitx}
\usepackage{multicol}
\usepackage{pifont}  
\usepackage{newunicodechar}
\newunicodechar{✖}{\ding{55}}
\usepackage{booktabs}   
\usepackage{pdflscape}  
\usepackage{ragged2e}

\usepackage{xcolor}
\usepackage[most]{tcolorbox}
\usepackage{pgfplots}
\pgfplotsset{compat=1.17}

\usepackage{booktabs}
\usepackage{float}                 
\usepackage{hyperref}              
\usepackage{url}
\usepackage{xcolor}                
\usepackage{lscape}                
\usepackage{caption}
\usepackage{subcaption}
\usepackage{etoolbox}
\usepackage{rotating}
\usepackage{multirow}
\usepackage{xcolor}
\usepackage{adjustbox}
\pgfplotsset{compat=1.17}

\AtBeginEnvironment{abstract}{\raggedright}
\AtBeginDocument{\RenewCommandCopy\qty\SI} 
\DeclareUnicodeCharacter{0301}{'} 

\begin{document}

\title{Quantum and classical approaches to the optimization of highway platooning: the two-vehicle matching problem
}

\hypersetup{
    pdfauthor   = {<First Author>, <Second Author>},
    pdftitle    = {Quantum- and Classical-Optimized Windbreaking-as-a-Service for Extended EV Range and CO₂ Savings},
    pdfkeywords = {quantum optimization, QUBO, electric vehicles, aerodynamic platooning, CO2 mitigation},
    colorlinks  = true,
    linkcolor   = teal,
    citecolor   = teal,
    urlcolor    = teal
}

\definecolor{Hungarian}{RGB}{79,129,189}      
\definecolor{MIQP}{RGB}{192,80,77}            
\definecolor{SA}{RGB}{155,187,89}             
\definecolor{Tabu}{RGB}{128,100,162}          
\definecolor{Leap}{RGB}{75,172,198}           
\definecolor{QA}{RGB}{247,150,70}             
\definecolor{QAOA}{RGB}{31,78,121}            

\titlerunning{Quantum \& classical approaches to the optimization of highway platooning}
%






\author{Chinonso Onah\inst{1,2}\orcidID{0000-0002-6296-533X} \and
\\Agneev Guin\inst{1,4}\orcidID{0000-0001-7065-2192} \and
\\Carsten Othmer\inst{1}\orcidID{0009-0009-3534-371X} \and
J.~A.~Montañez-Barrera\inst{3}\orcidID{0000-0002-8103-4581} \and
\\Kristel Michielsen\inst{2,3}\orcidID{0000-0003-1444-4262}}
\authorrunning{C. Onah et al.}
%
\institute{Volkswagen AG, Berliner Ring 2, Wolfsburg 38440, Germany \\
\email{\{chinonso.calistus.onah, agneev.guin, carsten.othmer\}@volkswagen.de} \and
Department of Physics, RWTH Aachen University, Germany \and
Jülich Supercomputing Centre, Forschungszentrum Jülich, Germany \\
\email{\{j.montanez-barrera,k.michielsen\}@fz-juelich.de} \and
Department of Network Engineering, BarcelonaTech (UPC), Spain}
\maketitle


\begin{abstract}
\justifying
Aerodynamic drag reduction on highways through vehicle platooning is a well-known concept that has not seen systematic uptake so far. Arguably because of high technological and legislative obstacles. As a low-tech entry solution to real multi-vehicle platooning, “Windbreaking-as-a-Service” (WaaS) was introduced in ~\cite{onah2025}.   Here we leverage the QUBO formulation to explore classical meta-heuristics such as simulated annealing, tabu search, and emergent quantum heuristics including quantum annealing and variants of the Quantum Approximate Optimization Algorithm (QAOA). These heuristic solvers do not guarantee optimality, yet they traverse the same high-order landscape in polynomial memory. They can be parallelized aggressively, and efficient classical processing can be employed to return \emph{only} valid schedules in certain hybrid workflows.  The present paper therefore positions QUBO  as the \emph{lingua franca} that allows heterogeneous classical, quantum and hybrid solvers to attack  the optimization of highway platooning. 
\end{abstract}




\section{Problem Formulation}
\label{sec:problem}

\subsection{Matching Problem Statement}

We study the two-vehicle matching problem on a single road segment formulated in ~\cite{onah2025}. 
A \emph{surfer} is a vehicle that seeks to follow (or be assigned to) a compatible lead vehicle, a \emph{breaker} for cooperative driving. The \emph{breaker} can accept at most one \emph{surfer} for such an assignment. 
The overall objective is to compute a one-to-one (or capacity-respecting) matching between surfers and breakers that minimizes a weighted cost capturing aerodynamic efficiency and timing/velocity compatibility. The a mathematical statement is the following: 

\medskip
Let the set of surfers be \(\mathcal S = \{ 1, \dots, S\}\) and the set of breakers be \(\mathcal B = \{ 1, \dots, B\}\).
We form a bipartite graph \(G(\mathcal S \cup \mathcal B , \mathcal E)\) where each edge \((s,b)\in \mathcal E\) indicates that surfer \(s \in \mathcal S\) can potentially be matched to breaker \(b \in \mathcal B\). For each possible edge \((s,b)\in \mathcal E\), we define a binary variable \(x_{s,b}\in\{0,1\}\) such that:
\[
x_{s,b} = 
\begin{cases}
1, & \text{if surfer $s$ is assigned to breaker $b$,}\\
0 & \text{otherwise.}
\end{cases}
\]

\noindent
The edge carries a weight $w_{s,b}$ reflecting the aerodynamic efficiency and timing compatibility of the pairing. It is defined as:
\begin{equation}
w_{s,b}
\;=\;
c_s\, V_b^2\, \bigl[\,1 - f(C_b - c_s)\bigr]
\;+\;
\lambda_1 \,\Delta_{s,b}^{T}
\;+\;
\lambda_2 \,\Delta_{s,b}^{V},
\label{eq:weight_formula}
\end{equation}
where \(\lambda_1, \lambda_2 \ge 0\) are scaling factors, and \(\Delta_{s,b}^{T}\) and \(\Delta_{s,b}^{V}\) capture time and velocity preferences of surfer \(s\) for breaker \(b\) as follows:
\begin{eqnarray}
\Delta_{s,b}^{T} &=&
\begin{cases} 
|t_s-T_b| & \text{if } |t_s-T_b| > \delta t_s/2, \\ 
0 & \text{otherwise,}
\end{cases}
\end{eqnarray}
and similarly for $\Delta_{s,b}^{V}$, with $\delta t_s$ and $\delta v_s$ being surfer-individual flexibility intervals for departure time and preferred travel speed, respectively.

\medskip
\noindent
The matching problem is therefore formalized in \cite{onah2025} as:
\begin{equation}
\min_{x_{s,b} \in \{0,1\}} \sum_{s \in S}\sum_{b \in B} w_{s,b}\,x_{s,b}
\label{eq:objective}
\end{equation}
subject to constraints:
\begin{equation}
\begin{aligned}
\sum_{b \in B} x_{s,b} &= 1,\quad \forall s \in S \quad (\text{each surfer matched exactly once})\\
\sum_{s \in S} x_{s,b} &= 1,\quad \forall b \in B \quad (\text{each breaker matched at most once})\;
\end{aligned}
\label{eq:constraints}
\end{equation}

\noindent
i.e. we require each surfer \(s \in \mathcal S\) to be matched to exactly one breaker, and each breaker \(b \in \mathcal B\) to accept exactly one surfer. We further assume $B=S$ in the rest of the paper.

\subsection{QUBO Form}

To exploit quantum and quantum-inspired solvers, we need to transform the problem into a QUBO problem. To that end, we add squared-penalty terms with a hyperparameter \(\lambda_3\) to quadratize the constraints and obtain the following unconstrained version:

\begin{equation}\label{eq:quad_penalties}
\lambda_3 \sum_{s\in \mathcal S} 
\Bigl(1 - \sum_{b\in \mathcal B} x_{s,b}\Bigr)^{2}
\;+\;
\lambda_3 \sum_{b\in \mathcal B}
\Bigl(1 - \sum_{s\in \mathcal S} x_{s,b}\Bigr)^{2}.
\end{equation}

This is added to Eq.~\eqref{eq:objective} to yield the QUBO problem. Namely:

\begin{equation}
\label{eq:final_qubo}
\min_{\{x_{s,b}\}} ( \sum_{(s,b)\in \mathcal E} w_{s,b}\,x_{s,b} \nonumber +\, \lambda_3 \sum_{s\in \mathcal S}\Bigl(1 - \sum_{b\in \mathcal B} x_{s,b}\Bigr)^2 \nonumber +\, \lambda_3 \sum_{b\in \mathcal B}\Bigl(1 - \sum_{s\in \mathcal S} x_{s,b}\Bigr)^2).
\end{equation}

With \(n:=S(=B)\), we can list all pairs \((s,b)\) in a vector:
\[
x \;=\; \bigl(
x_{1,1},\, x_{1,2},\,\dots,\, x_{1,n},\,
x_{2,1},\,\dots,\, x_{n,n}
\bigr)^T.
\]

Eq.~\eqref{eq:final_qubo} becomes the standard form as mentioned in \cite{onah2025}:

\begin{equation}
\min \, x^T Q x
\label{eq: qubo}
\end{equation}

\begin{align}
\text{with} \quad Q_{ee} = w_{s,b} + \lambda_1 \Delta_{s,b}^{T} + \lambda_2 \Delta_{s,b}^{V}- 2 \lambda_3
\notag \\
\text{and} \quad Q_{ee'} = 
\begin{cases} 
\lambda_3 & \text{if } e \text{ and } e' \text{ share a node} \\ 
0 & \text{otherwise.}
\end{cases}\notag
\end{align}

\noindent
This QUBO matrix has an interesting tensor structure. To see that, let $Q \in \mathbb{R}^{n^2 \times n^2}$ and

\begin{equation}
Q = \text{diag}_{n^2} \left( w_{s,b} - 4 \lambda_3 \right)
+ \lambda_3 \left( I_N \otimes J_N + J_N \otimes I_N \right)
\label{eq:qubo_matrix}
\end{equation}

\begin{equation}
\begin{aligned}
\text{where } I_N: N \times N \text{ identity matrix}  \notag \\
\text{and }J_N: N \times N \text{ ``1''-matrix} =
\begin{pmatrix}
1 & 1 & \cdots & 1 \\
1 & 1 & \cdots & 1 \\
\vdots & \vdots & \ddots & \vdots \\
1 & 1 & \cdots & 1
\end{pmatrix}
\end{aligned}
\end{equation}

\noindent
This structure is closely related to the fact that valid solutions can be translated into permutation matrices. This is a property that can be leveraged to improve the quantum solution both in speed and accuracy. (Refer Sec. \ref{sec:ce-qaoa} and ~\cite{onahce}).

\subsubsection{Constraints and Energy Landscape Analysis}

We seek a small enough penalty weight \(\lambda_3\) that guarantees all invalid one‐hot assignments lie strictly above all valid ones in the QUBO energy landscape. To derive this, let the matrix representation of the problem \(C = [c_{ij}]\in\mathbb{R}^{n\times n}\).  The maximum possible cost‐saving from swapping any one assignment is
\[
c_{\max} - c_{\min},
\]
and there are \(n\) rows (or columns) whose one‐hot constraints could be violated.  Hence any invalid bitstring can gain at most \(n\,(c_{\max}-c_{\min})\) in the cost term.  By choosing
\[
\lambda_3 > n\,(c_{\max}-c_{\min}),
\]
we ensure that the penalty for violating even one one‐hot constraint outweighs \emph{all} possible cost savings from invalid assignments.  Multiplying by a \emph{safety} factor \(>1\) yields a strict separation without unduly inflating the penalty.

\usetikzlibrary{arrows.meta,positioning,calc}

\begin{figure}[h]
\centering
\begin{tikzpicture}[
  >=Latex,
  feas/.style={blue!70!black, fill=blue!18},
  infeas/.style={red!70!black, fill=red!18},
  lab/.style={font=\small},
  title/.style={font=\small\bfseries}
]

\begin{scope}[xshift=0cm]
  \draw[->] (0,0) -- (0,4.3) node[above,lab]{energy $E(\mathbf{x})$};
  \draw[-] (-0.1,0) -- (3.4,0);

  \node[title,anchor=west] at (0.2,4.15) {$\lambda_3$ too small};

  \draw[feas,rounded corners] (0.55,1.60) rectangle (1.55,2.45);
  \node[lab,blue!70!black] at (1.05,2.70) {feasible};

  \draw[infeas,rounded corners] (2.05,1.10) rectangle (3.05,2.90);
  \node[lab,red!70!black] at (2.55,3.15) {infeasible};

  \fill[red!75!black] (2.55,1.25) circle (1.8pt);
  \draw[->,red!75!black] (2.55,1.40) -- (1.55,1.62);
  \node[lab,align=left,red!75!black,anchor=west] at (0.2,0.55)
    {infeasible strings can\\ drop below feasible minima\\ (``cheating'')};

  \foreach \y in {1.75,2.00,2.20,2.35} {
    \fill[blue!70!black] (1.05,\y) circle (1.5pt);
  }
  \foreach \y in {1.55,2.10,2.55,2.75} {
    \fill[red!70!black] (2.55,\y) circle (1.5pt);
  }

  \node[lab,anchor=west] at (0.2,3.55)
    {$E(\mathbf{x}) = E_{\text{cost}}(\mathbf{x}) + \lambda_3 P(\mathbf{x})$};
\end{scope}

\begin{scope}[xshift=7.0cm]
  \draw[->] (0,0) -- (0,4.3) node[above,lab]{energy $E(\mathbf{x})$};
  \draw[-] (-0.1,0) -- (3.4,0);

  \node[title,anchor=west] at (0.2,4.15) {$\lambda_3$ very large};

  \draw[feas,rounded corners] (0.55,1.95) rectangle (1.55,2.10);
  \node[lab,blue!70!black] at (1.05,2.35) {feasible};

  \draw[infeas,rounded corners] (2.05,3.05) rectangle (3.05,3.85);
  \node[lab,red!70!black] at (2.55,4.05) {infeasible};

  \draw[<->,thick] (1.80,2.10) -- (1.80,3.05);
  \node[lab,anchor=west] at (1.92,2.58) {penalty gap};

  \draw[->,blue!70!black] (1.55,2.02) -- (2.65,1.55);
  \node[lab,align=left,blue!70!black,anchor=west] at (0.2,0.55)
    {feasible energies nearly equal\\ (flat feasible plateau)\\ $\Rightarrow$ weak cost discrimination};

  \foreach \y in {1.975,2.01,2.05} {
    \fill[blue!70!black] (1.05,\y) circle (1.5pt);
  }
  \foreach \y in {3.15,3.35,3.60,3.75} {
    \fill[red!70!black] (2.55,\y) circle (1.5pt);
  }

  \node[lab,anchor=west] at (0.2,3.55)
    {$E(\mathbf{x}) \approx \lambda_3 P(\mathbf{x})$ on infeasible set};
\end{scope}

\end{tikzpicture}
\caption{Schematic energy landscape under two penalty-weight regimes. Left:
if $\lambda_3$ is too small, infeasible configurations can ``cheat'' by lowering
$E_{\text{cost}}$ enough to overcome penalties. Right: if $\lambda_3$ is very large,
infeasible states are pushed far above the feasible manifold, but feasible energies
compress into an almost flat plateau, reducing cost discrimination among valid
matchings.}
\label{fig:penalty-regimes}
\end{figure}
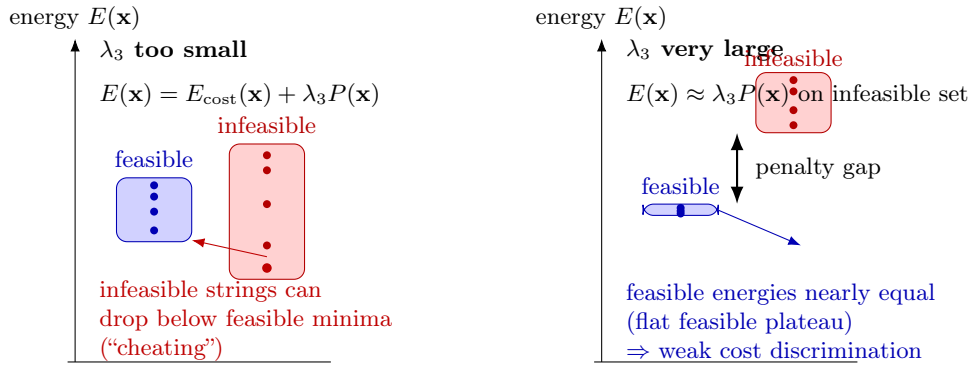



\medskip
\noindent
As shown in Fig. \ref{fig:penalty-regimes}, when $\lambda_3$ is too small compared to the spread of~$c_{i,j}$, then the
sampler can \emph{reduce} $E_{\mathrm{cost}}$ by picking invalid
assignments (e.g.\ two ones in the same row), because the small
penalty is outweighed by cost savings.  This produces noisy, infeasible
solutions that cheat the one-hot rule in Eq. \eqref{eq:constraints}. 

\subsubsection{Mapping the QUBO to a Quantum Hamiltonian}
To solve an optimization problem on a quantum computer, its QUBO formulation must be mapped to a quantum Hamiltonian~\cite{Lucas_2014}. This is derived in the standard way by expressing a symmetric QUBO matrix in terms of its components and transforming the upper triangular part into an Ising Hamiltonian. In this formalism, each qubit in the computational basis corresponds to a binary variable and hence an edge \((s,b)\) in the bipartite graph picture. 
 Consider the following generic QUBO matrix:

\begin{equation}
x^T Q x = \sum_{i,j} Q_{ij} x_i x_j
\end{equation}

where \( x_i \in \{0, 1\} \) are binary variables, and \( Q_{ij} \) is the matrix defining the quadratic coefficients. Each binary variable \( x_i \) is mapped to a qubit using the relation:

\begin{equation}
x_i = \frac{1}{2} (I - Z_i)
\end{equation}

where \( Z_i \) is the Pauli-Z operator acting on the \( i \)-th qubit, and \( I \) is the identity operator, effectively transforming the binary variable \( x_i \) into a Pauli-Z operator. See Ref~\cite{Lucas_2014} for details. The full Hamiltonian becomes

\begin{equation}
\label{eq:quboe}
H = \sum_{i} \frac{Q_{ii}}{2} (I - Z_i) + \sum_{i < j} \frac{Q_{ij}}{4} \left( I - Z_i - Z_j + Z_i Z_j \right) .
\end{equation}

The constant identity terms do not influence the optimization landscape of the QUBO problem and can be omitted.

\subsection{Performance metrics}\label{sec:metrics}

Because heuristic samplers return \emph{distributions} rather than proofs of optimality, a single scalar such as the “best cost” gives an incomplete—and occasionally misleading—picture of performance.  A solver call returns \(N\) raw samples (``reads'') of bitstrings \(\mathbf x_k\in\{0,1\}^{N_{\text{vars}}}\)\footnote{$N_{\text{vars}}$ is the number of binary variables in the problem and thus the length of the bitstring}, their QUBO energies
\(
E_k \equiv E(\mathbf x_k)
\),
and quadratic penalties
\(
P(\mathbf x_k)
\)
which vanish only for feasible matchings (Refer Sec. ~\ref{sec:problem}).  We denote by \(E_\star\) the exact ground-state energy obtained from the Hungarian algorithm and the corresponding MIQP (mixed-integer quadratic program) baseline (Refer Sec. ~\ref{sec:mip}). Following the benchmarking practice of Refs.~\cite{ronnow2014speedup,vert2021benchmarking}, we log an additional battery of complementary metrics for different solvers and aggregate them for further statistical analyses. The following metrics are considered. 

\begin{enumerate}[leftmargin=1.7em]

\item \textbf{Best energy:}
\[
  E_{\text{best}} \;=\; \min_{k} E_k.
\]
This is the smallest energy observed in a batch and is the headline quality indicator: if \(E_{\text{best}} = E_\star\), the sampler has, at least once, hit a global optimum (up to degeneracy).  Because this metric is sensitive to rare outliers, we always report it alongside aggregate statistics.

\item \textbf{Mean energy:}
\[
  \langle E\rangle \;=\; \frac{1}{N}\sum_{k=1}^{N} E_k.
\]
The mean energy smooths out incidental hits and captures the typical quality an end user would obtain by accepting the first sample supplied by the solver.

\item \textbf{Optimality gaps:}
When an exact optimum \(E_\star\) is available (via the Hungarian or MIQP baselines), we quantify absolute solution quality through relative gaps
\[
  \Delta_{\text{best}}
  \;=\;
  \frac{E_{\text{best}} - E_\star}{\lvert E_\star\rvert},
  \qquad
  \Delta_{\text{mean}}
  \;=\;
  \frac{\langle E\rangle - E_\star}{\lvert E_\star\rvert}.
\]

\item \textbf{Energy variance:}
\[
  \sigma^{2}_E
  \;=\;
  \frac{1}{N}\sum_{k=1}^{N}\bigl(E_k-\langle E\rangle\bigr)^2.
\]
High \(\sigma^2_E\) indicates broad exploration of the energy landscape; low variance indicates a sharply peaked output distribution, which may reflect successful concentration around good minima—or, conversely, premature convergence to a poor local basin.  We therefore interpret this metric in conjunction with the gaps and success probability.

\item \textbf{Single-shot success probability (ground-state probability):}
The success probability is defined as 
\begin{equation}
  p_{\mathrm{succ}}
  \;:=\;
  \Pr_{\mathbf x\sim \pi}
  \bigl[E(\mathbf x) = E_\star \ \wedge\ P(\mathbf x)=0\bigr],
  \label{eq:psucc-def}
\end{equation}
Empirically, we estimate it from a batch of
\(N\) samples \(\{\mathbf x_k\}_{k=1}^N\) as
\[
  \hat p_{\mathrm{succ}}
  \;=\;
  \frac{
    \#\bigl\{k :
      E(\mathbf x_k) = E_\star \ \wedge\ P(\mathbf x_k)=0
    \bigr\}
  }{N},
\]
i.e.\ the fraction of all samples that are feasible and achieve the ground-state energy. If the ground state of the problem Hamiltonian is degenerate, two or more bitstrings that correspond to different matching outcomes can attain this energy value. This scenario is further discussed in Sec. \ref{sec:cost-to-impact} where one readily observes different assignments that match the same MIQP energy leading to different energy-saving outcomes compared to the MIQP and Hungarian outcomes.

\item \textbf{Samples-to-solution (\textbf{STS})}
For independent samples, the expected number of draws needed to observe a single optimum
is the mean of a geometric distribution,
\begin{equation}
  \mathrm{STS}
  \;:=\;
  \frac{1}{p_{\mathrm{succ}}},
  \label{eq:sts-def}
\end{equation}
as in Eq.~\eqref{eq:sts-def}. In practice we plug in the empirical estimate
\(\hat p_{\mathrm{succ}}\). STS is backend-agnostic: it depends only on the sampler's
quality, not on device speed.

\end{enumerate}


\subsection{Benchmark Dataset}
\label{sec:dataset}

We publish a small companion dataset containing ten single–segment matching instances with $n\in\{3,\dots,12\}$ surfer–breaker pairs. For each $n$ we
provide two \texttt{NumPy} arrays,
\texttt{n-vehicles-breakers.npy} and \texttt{n-vehicles-surfers.npy}. The
breaker array stores, per vehicle, its class label, cruising velocity, and departure time, while the surfer array stores the surfer’s class, preferred
velocity, departure time, and individual flexibility intervals for speed and
departure time. These instances are used to construct the edge weights
$w_{s,b}$ and the corresponding QUBO matrices $Q$ for all experiments reported
in this work. A python script for manipulating the data is also provided. The dataset is versioned and publicly available at
\url{https://doi.org/10.5281/zenodo.17768136}\cite{onah2025data}. The dataset is used for the tables and figures in Sec. ~\ref{sec:methods} and \ref{sec:cost-to-impact}.

\section{Quantum-(Inspired) Optimization Methods}\label{sec:methods}

All seven solvers in our ``solver zoo'' act on the same combinatorial object: either the linear assignment problem of Eqs.~\eqref{eq:objective}–\eqref{eq:constraints} or its QUBO reformulation in Eqs.~\eqref{eq:final_qubo}-\eqref{eq:qubo_matrix}.  For the latter, it is convenient to write
\begin{equation}
  E(\mathbf x)
  \;=\;
  \mathbf x^{\mathsf T} Q\,\mathbf x,
  \qquad
  \mathbf x \in \{0,1\}^{N},
  \quad N = n^2,
  \label{eq:qubo-energy-compact}
\end{equation}
with ground-state energy \(
  E_\star = \min_{\mathbf x} E(\mathbf x).
\) Each heuristic solver defines (implicitly) a probability distribution
\(
\pi(\mathbf x)
\)
over bitstrings.  

\begin{figure}[h]
\centering
\begin{tikzpicture}[
    node distance = 0.25cm and 2.9cm,
    solver/.style  ={
      draw, rounded corners, thick,
      minimum width=3.0cm, minimum height=1.4cm,
      font=\sffamily\large, fill=gray!15,
      text=white
    },
    dataset/.style ={
      draw, thick, fill=yellow!25,
      rounded corners, align=center,
      minimum width=3.2cm, minimum height=1.3cm,
      font=\sffamily\large
    },
    metricbox/.style={
      draw, thick, fill=blue!8,
      rounded corners, align=left,
      inner sep=6pt,
      minimum width=4.0cm,
      font=\sffamily\large
    },
    arr/.style     ={thick, -{Latex[length=2mm]}}
]

  \node[solver, fill=Hungarian]            (hun)  {Hungarian};
  \node[solver, fill=MIQP, below=of hun]  (miqp) {MIQP};
  \node[solver, fill=SA, below=of miqp] (sa)   {Simulated Annealing};
  \node[solver, fill=Tabu, below=of sa]   (tabu) {Tabu search};
  \node[solver, fill=Leap, below=of tabu] (leap) {Leap Hybrid};
  \node[solver, fill=QA, below=of leap]     (qa)   {Quantum Annealing};
  \node[solver, fill=QAOA, below=of qa] (qaoa) {QAOA};

  \coordinate (mid) at ($(hun)!.5!(qaoa)$);

  \node[dataset, left=of mid] (data) {Fixed\\problems};

  \node[metricbox, right=of mid] (metrics) {%
    Optimality    Gap (\%)\\
     Time to Solution (TTS)\\
     Sample to Solution (STS)
  };



    \draw[arr] (data.east) -- ++(0.5,0) |- (hun.west);
    \draw[arr] (data.east) -- ++(0.5,0) |- (miqp.west);
    \draw[arr] (data.east) -- ++(0.5,0) |- (sa.west);
    \draw[arr] (data.east) -- ++(0.5,0) |- (tabu.west);
    \draw[arr] (data.east) -- ++(0.5,0) |- (leap.west);
    \draw[arr] (data.east) -- ++(0.5,0) |- (qa.west);
    \draw[arr] (data.east) -- ++(0.5,0) |- (qaoa.west);
    
    \draw[arr] (hun.east) -| ([xshift=-0.5cm]metrics.west);
    \draw[arr] (miqp.east) -| ([xshift=-0.5cm]metrics.west);
    \draw[arr] (sa.east) -| ([xshift=-0.5cm]metrics.west);
    \draw[arr] (tabu.east) -| ([xshift=-0.5cm]metrics.west);
    \draw[arr] (leap.east) -| ([xshift=-0.5cm]metrics.west);
    \draw[arr] (qa.east) -| ([xshift=-0.5cm]metrics.west);
    \draw[arr] (qaoa.east) -| ([xshift=-0.5cm]metrics.west);
    \draw[arr] (tabu.east) -- ++(0.5,0) |- (metrics.west);

\end{tikzpicture}
\caption{“Solver-zoo” benchmark. Every solver is run on the \textbf{same fixed problem sets} (yellow), and all outputs are funneled into one \textbf{common metric panel} (purple) for direct comparison.}
\label{fig:solver-zoo}
\end{figure}
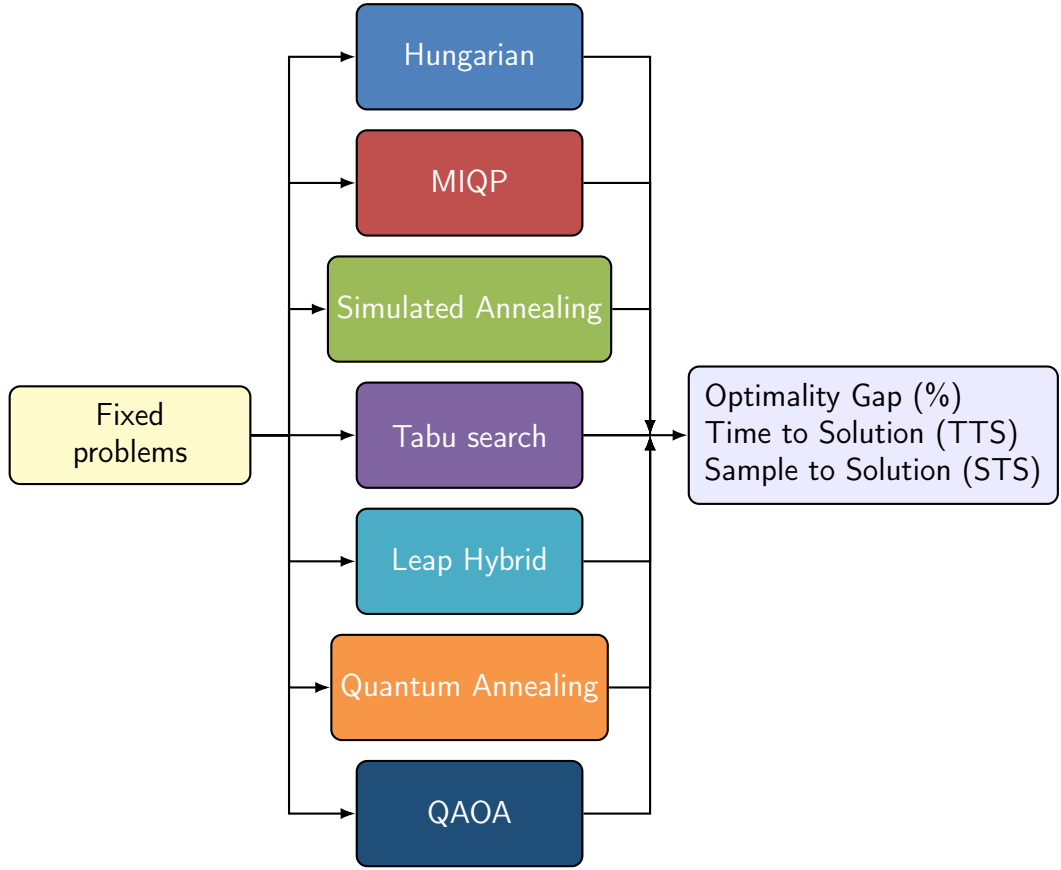

\subsection{Hungarian assignment algorithm: exact polynomial baseline}
\label{sec:hungarian}

The Hungarian algorithm solves the linear assignment problem
\begin{align}
  \min_{x_{ij}} \quad
    & \sum_{i=1}^{n}\sum_{j=1}^{n} c_{ij} x_{ij},
    \label{eq:hungarian-objective}
  \\[0.5ex]
  \text{s.t.}\quad
    & \sum_{j=1}^{n} x_{ij} = 1, \quad i=1,\dots,n,
 \notag \\
    & \sum_{i=1}^{n} x_{ij} = 1, \quad j=1,\dots,n,
\notag \\
    & x_{ij} \in \{0,1\},
    \label{eq:hungarian-binary}
\notag 
\end{align}
in strongly polynomial time $\mathcal O(n^3)$~\cite{kuhn1955hungarian}.  In our setting, $c_{ij}$ corresponds to the edge weights $w_{s,b}$ on a single road segment, i.e.\ the cost term of Eq.~\eqref{eq:objective} without quadratic penalties.  Because all feasible matchings satisfy the one-hot constraints, the minimizer of Eq.~\eqref{eq:hungarian-objective} coincides with the ground state of the QUBO energy restricted to the feasible subspace.

We therefore treat the Hungarian method as our \emph{exact polynomial baseline}: it delivers the provably optimal two-vehicle matching on each road segment, and its optimal value $E_{\mathrm{Hung}}$ serves as the reference against which we measure optimality gaps of all heuristic solvers,
\begin{equation}
  \Delta_\text{best}\;=\;
  \frac{E_{\mathrm{solver}} - E_{\mathrm{Hung}}}{\lvert E_{\mathrm{Hung}}\rvert}\times 100\%.
  \label{eq:gap-def}
\end{equation}

\subsubsection*{Hungarian baseline in the metrics language:}

The Hungarian algorithm is an \emph{exact} polynomial-time solver for the linear assignment problem.  On each instance it returns a unique optimal matching $\mathbf x_{\mathrm{Hung}}$ that minimizes the linear cost in Eq. ~\eqref{eq:hungarian-objective}.  When we embed this matching into the QUBO representation of Eq.~\eqref{eq:qubo-energy-compact} via the row-major bitstring
\(
  \mathbf x_{\mathrm{Hung}} \in \{0,1\}^{n^2},
\)
its QUBO energy
\(
  E_{\mathrm{Hung}} = E(\mathbf x_{\mathrm{Hung}})
\)
coincides with the ground-state energy $E_\star$ restricted to the feasible subspace.\footnote{Modulo the rescaling factors used to normalize the Ising Hamiltonian in Sec. ~\ref{sec:qa}.} As per the performance metrics of Sec. ~\ref{sec:metrics}, this implies
\[
  p_{\mathrm{succ}} = 1,
  \qquad
  \mathrm{STS} = \frac{1}{p_{\mathrm{succ}}} = 1.
\]
Every call to the Hungarian solver returns a feasible ground state, so the best and mean energies coincide,
\[
  E_{\text{best}} = \langle E\rangle = E_{\mathrm{Hung}},
  \qquad
  \sigma_E^2 = 0,
\]
and the optimality gaps $\Delta_{\text{best}}$ and $\Delta_{\text{mean}}$ also vanish by construction.

\begin{figure}[p]
  \centering

  \begin{subfigure}[t]{0.5\textwidth}
    \centering
    \includegraphics[width=\linewidth]{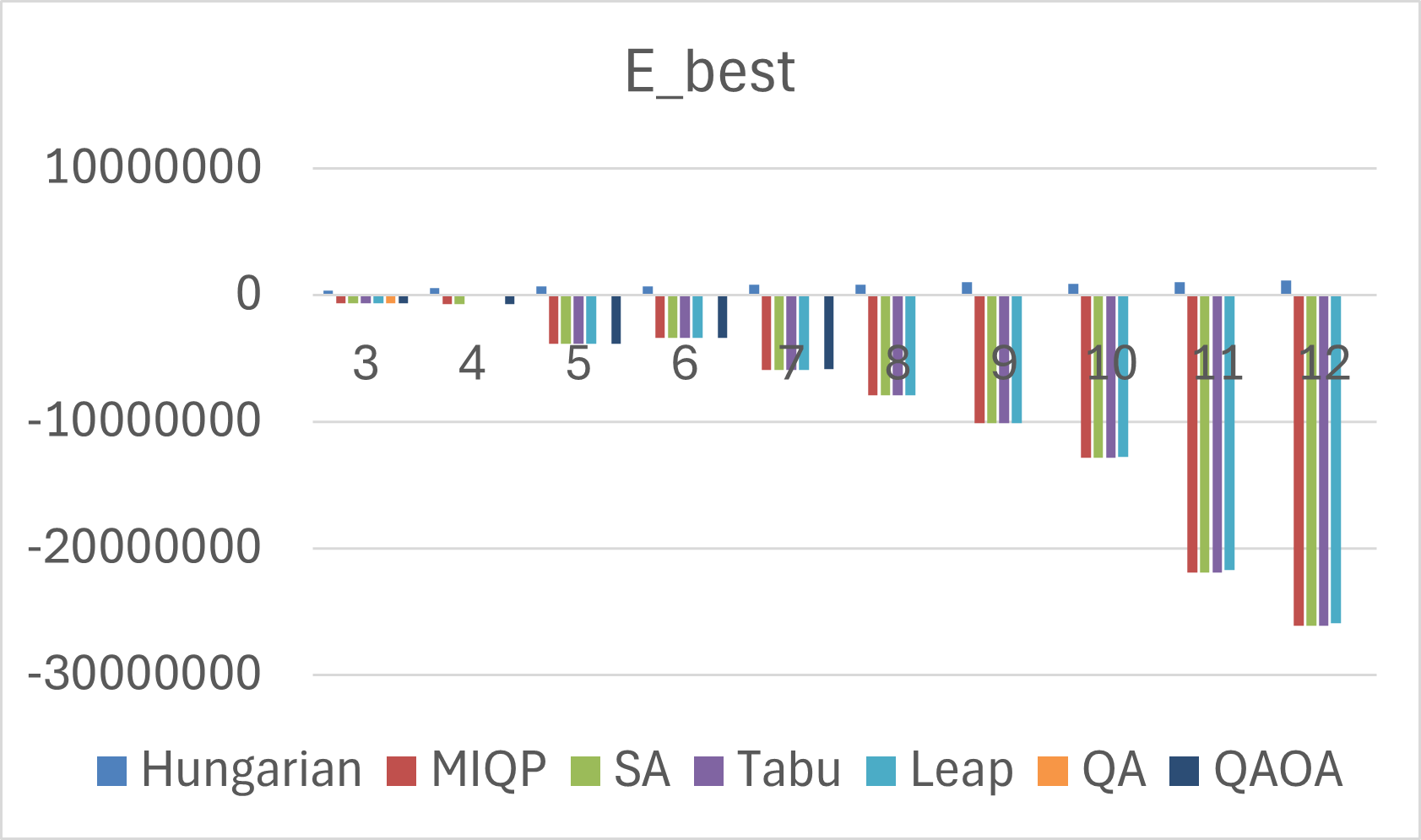}
    \caption{(Best energy)}
    \label{fig:metrics_1}
  \end{subfigure}\hfill
  \begin{subfigure}[t]{0.5\textwidth}
    \centering
    \includegraphics[width=\linewidth]{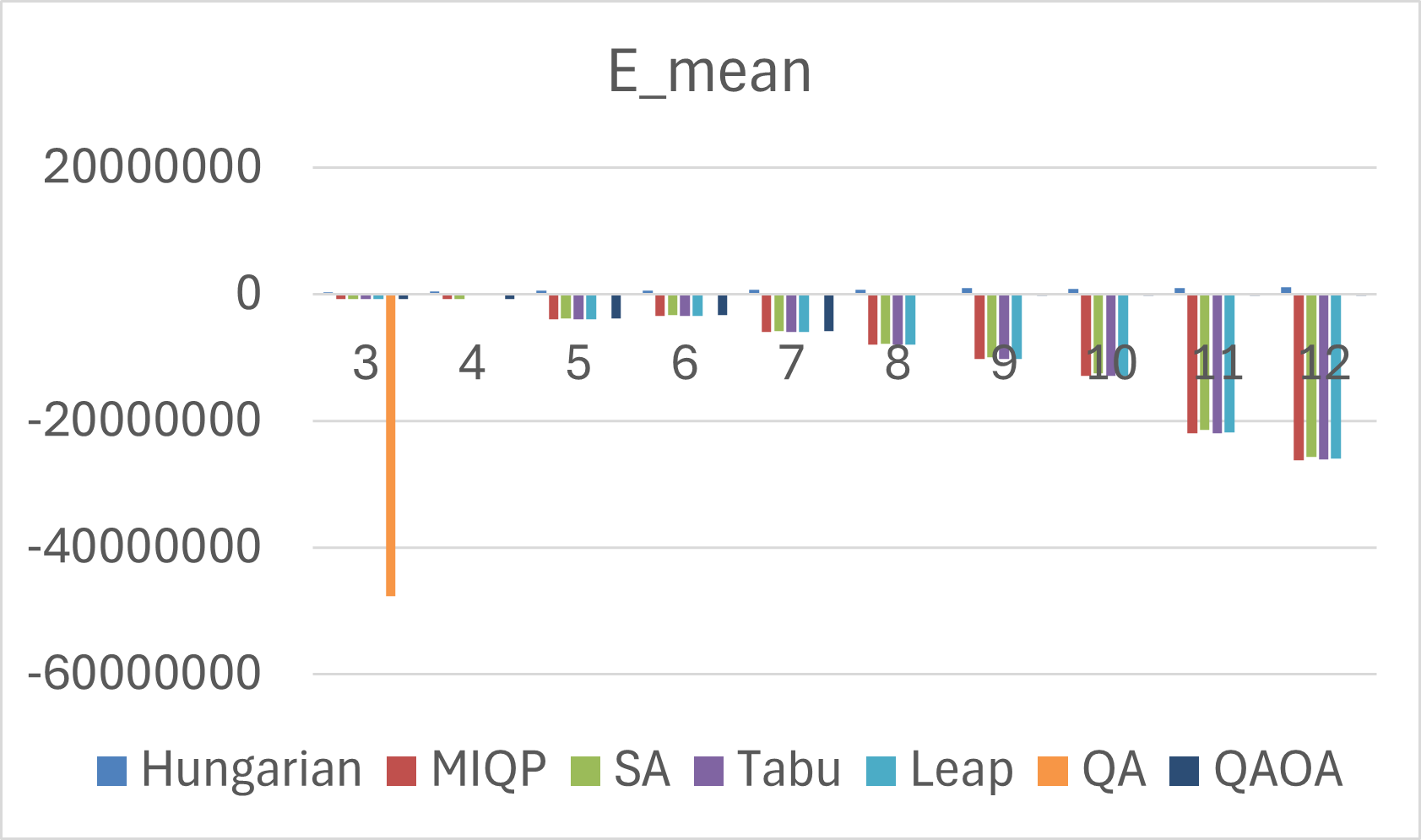}
    \caption{(Mean energy)}
    \label{fig:metrics_2}
  \end{subfigure}

  \vspace{0.6em}

  \begin{subfigure}[t]{0.5\textwidth}
    \centering
    \includegraphics[width=\linewidth]{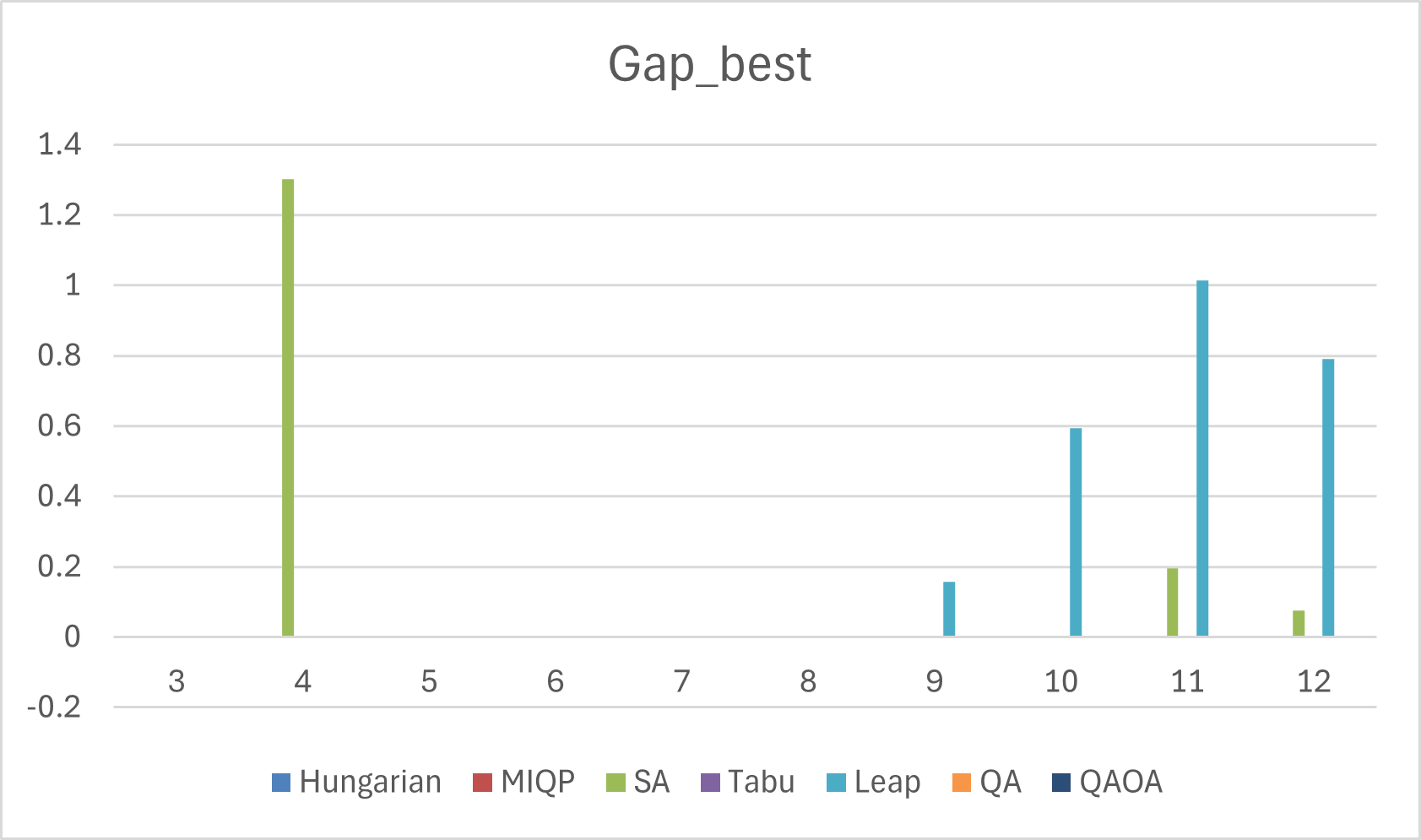}
    \caption{(Energy Gap (\%))}
    \label{fig:metrics_3}
  \end{subfigure}\hfill
  \begin{subfigure}[t]{0.5\textwidth}
    \centering
    \includegraphics[width=\linewidth]{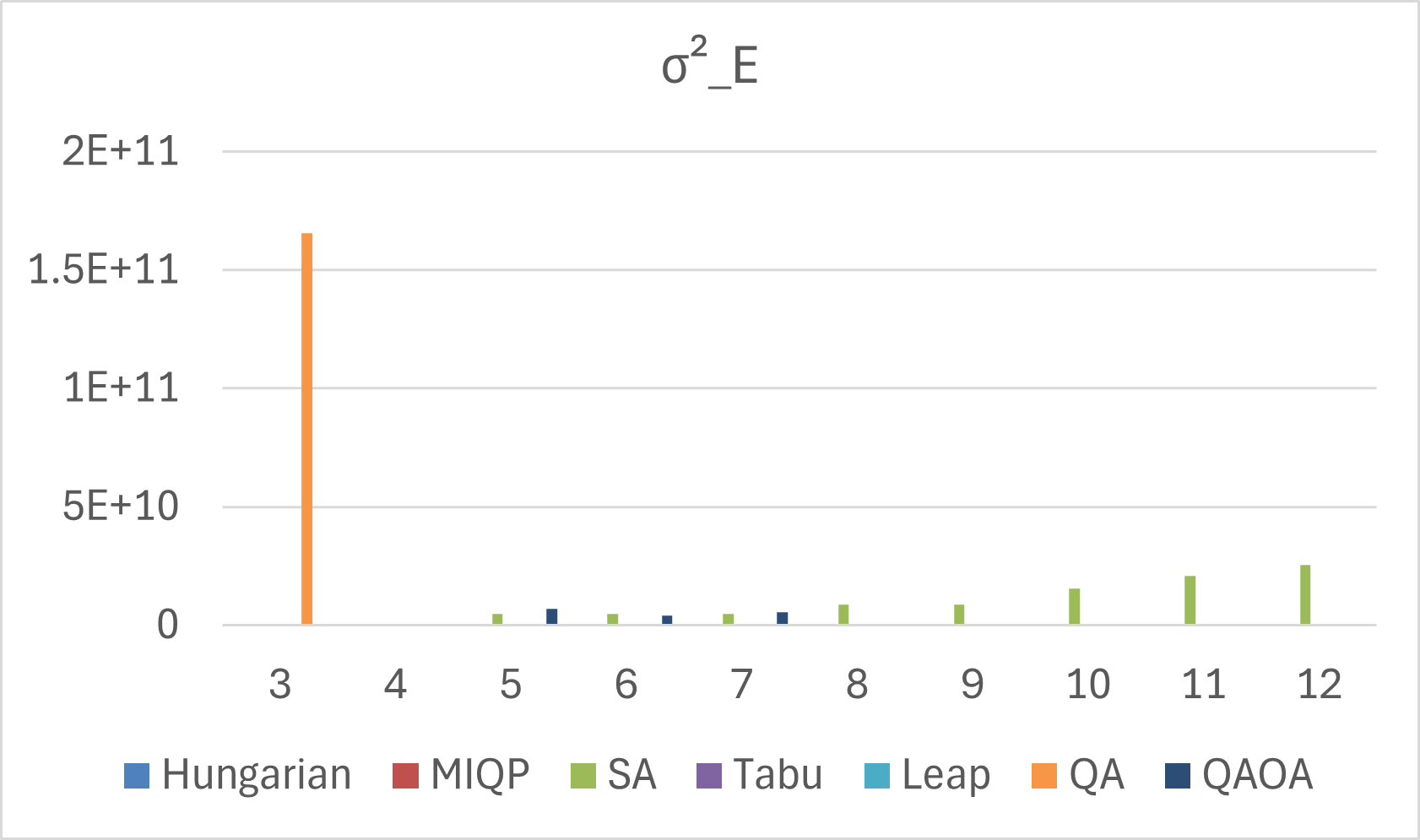}
    \caption{(Energy Variance)}
    \label{fig:metrics_4}
  \end{subfigure}

  \vspace{0.6em}

  \begin{subfigure}[t]{0.5\textwidth}
    \centering
    \includegraphics[width=\linewidth]{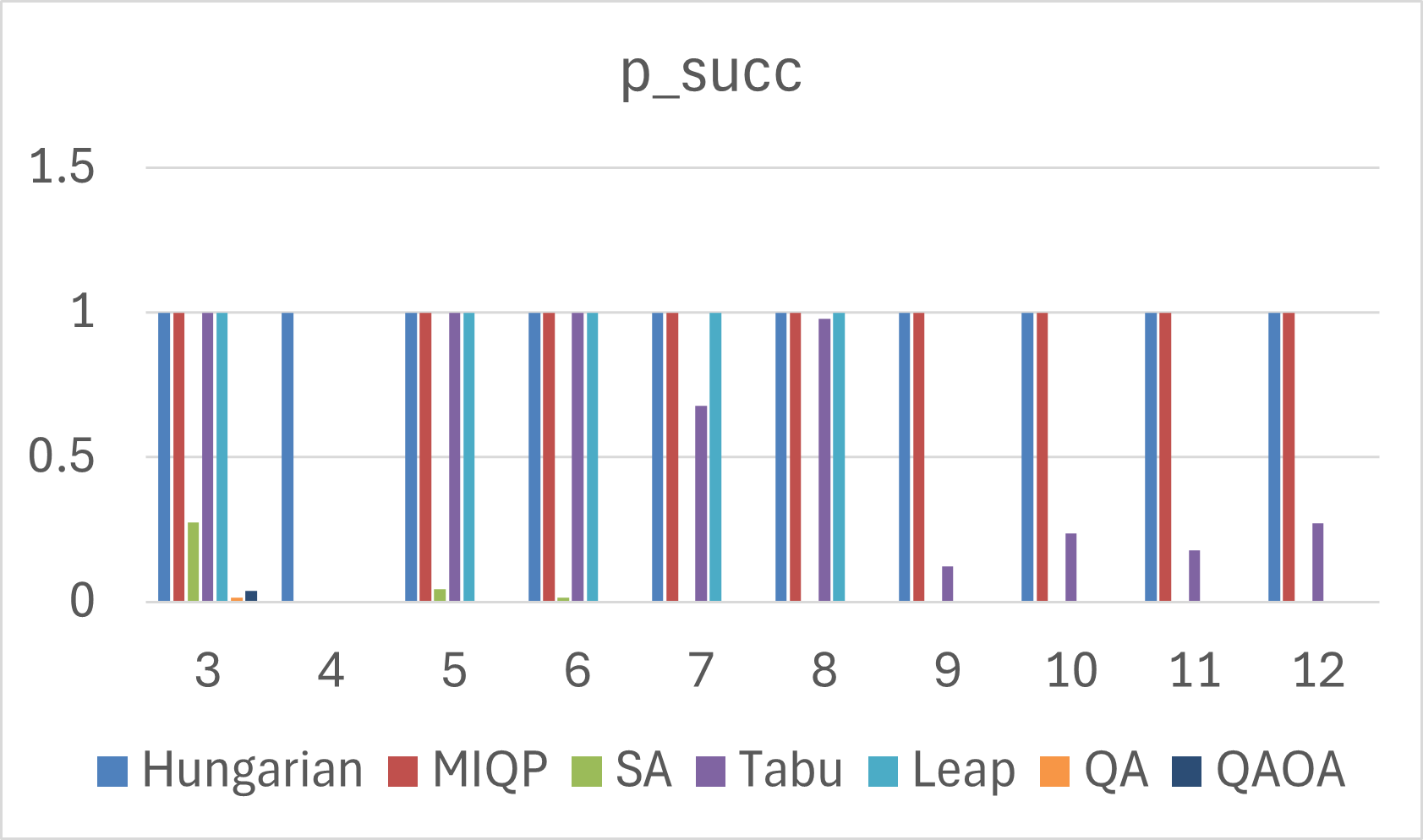}
    \caption{(Success probability)}
    \label{fig:metrics_5}
  \end{subfigure}\hfill
  \begin{subfigure}[t]{0.5\textwidth}
    \centering
    \includegraphics[width=\linewidth]{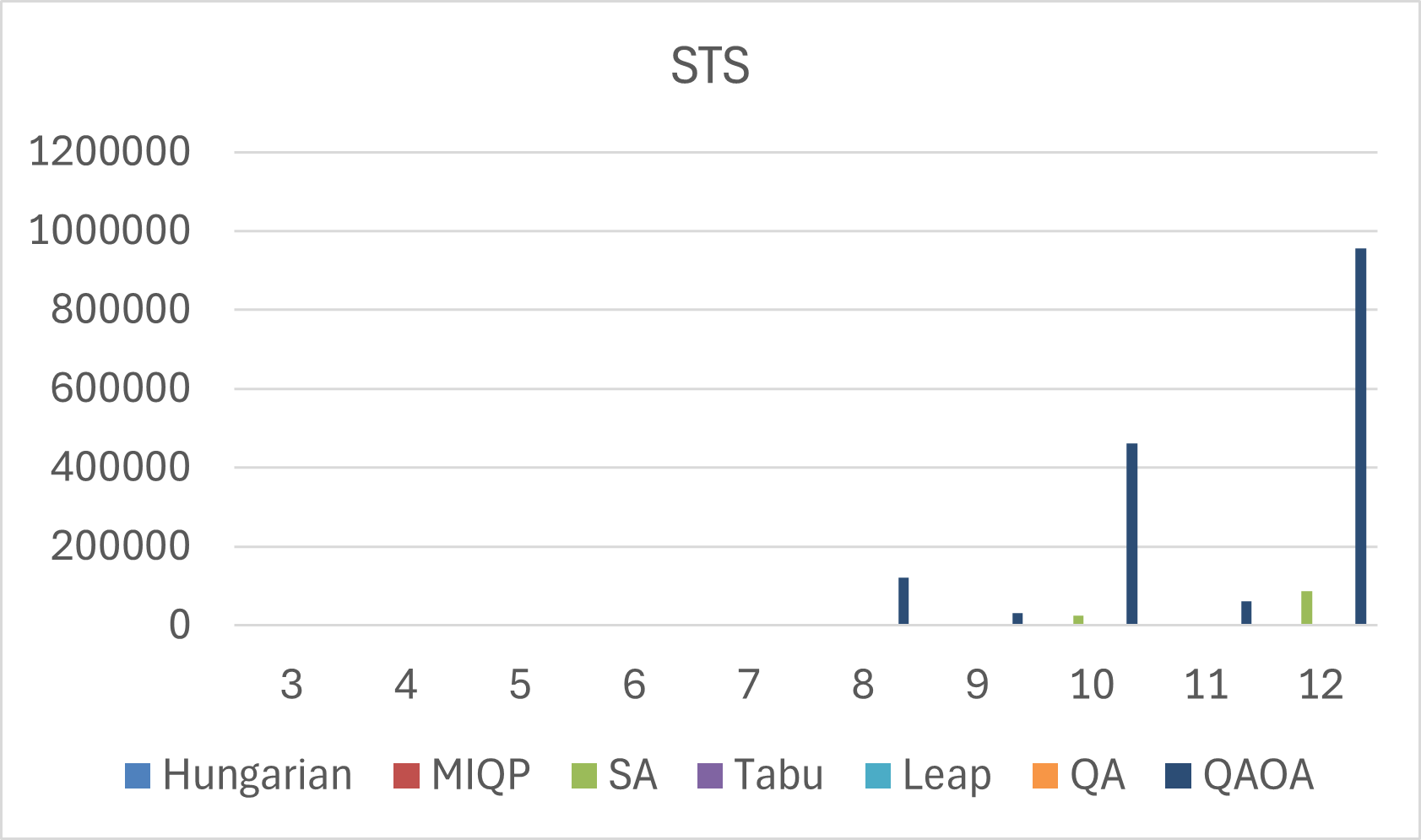}
    \caption{(Sample-to-solution)}
    \label{fig:metrics_6}
  \end{subfigure}

  \vspace{0.6em}


  \caption{Summary of solver performance metrics for 10 problem instances in the benchmark suite in Sec. \ref{sec:dataset}. For a more detailed view of the benchmark results underlying these plots and Figs.~\ref{fig:eta_threeplots}, we refer the reader to the Supplementary Material App. \ref{app:app}. It provides the full benchmarking tables for all instance sizes, including the raw and derived quantities used in the plots. These supplementary tables provide a reproducible reference for comparing the classical, quantum, and hybrid solvers.}
  \label{fig:metrics_gallery}
\end{figure}


\subsection{Gurobi MIQP solver}\label{sec:mip}

To obtain an exact baseline at the QUBO level, we also formulate the problem as a mixed-integer quadratic program (MIQP).  Introducing a binary decision vector
\(
\mathbf z \in \{0,1\}^{N}
\)
that coincides with $\mathbf x$, the MIQP reads
\begin{align}
  \min_{\mathbf z} \quad
    & \frac{1}{2}\,\mathbf z^{\mathsf T} H\,\mathbf z
      + \mathbf f^{\mathsf T}\mathbf z,
      \label{eq:miqp-obj}
  \\
  \text{s.t.}\quad
    & A\mathbf z = \mathbf b,
      \label{eq:miqp-constraints}
 \notag \\
    & \mathbf z \in \{0,1\}^{N},
\notag
\end{align}
where $H$ and $\mathbf f$ are constructed from the QUBO matrix $Q$ (up to a constant shift) and $A\mathbf z=\mathbf b$ encode the one-hot constraints explicitly instead of via penalties.  In practice we use the Gurobi Optimizer in MIQP mode~\cite{gurobi2024}, which combines presolve reductions, cutting planes, heuristics, and branch-and-bound/branch-and-cut tree search to obtain provably optimal solutions or certified bounds on $E_\star$.

For the problem sizes considered here, MIQP remains tractable and yields the optimal energy $E_{\mathrm{MIQP}}$, a certified lower bound $E_{\mathrm{LB}}$ and the corresponding \emph{MIQP gap}. We verify the optimality by matching the corresponding assignment cost with $E_{\mathrm{Hung}}$.
The \emph{MIQP gap} is the relative optimality gap reported by Gurobi \cite{gurobi2024}, and is defined by
\begin{align}
  \text{MIQP gap} = \frac{|E_{\text{best}} - E_{\text{LB}}|}{|E_{\text{best}}|}
  \notag
\end{align}
  
  MIQP thus provides a second, independent correctness oracle for the heuristic solvers and is consistent with recent work on quantum-ready routing formulations~\cite{harwood2021formulating}. The MIQP provided the necessary baseline  for benchmarking results reported in Figs. \ref{fig:metrics_gallery} and \ref{fig:eta_threeplots}. 

\noindent



\subsection{Simulated annealing (SA): thermodynamic meta-heuristic}

Simulated annealing (SA) is a stochastic local-search heuristic inspired by thermodynamic cooling~\cite{kirkpatrick1983optimization,geman1984annealing}.  On a given QUBO energy $E(\mathbf x)$ we define a single-spin-flip Markov chain over $\{0,1\}^N$ at temperature $T>0$.  Starting from an initial configuration $\mathbf x^{(0)}$, one iteration proposes a neighbor $\mathbf x'$ (e.g.\ by flipping a randomly chosen bit) and accepts it with Metropolis probability
\begin{equation}
  p_{\mathrm{acc}}(\mathbf x \to \mathbf x'; T)
  \;=\;
  \min\Bigl\{
    1,\;
    \exp\bigl[-\Delta E / T\bigr]
  \Bigr\},
\qquad
\Delta E = E(\mathbf x') - E(\mathbf x).
  \label{eq:sa-metropolis}
\end{equation}
At fixed temperature $T$, the chain has the Boltzmann distribution
\(
\pi_T(\mathbf x) \propto e^{-E(\mathbf x)/T}
\)
as stationary measure under mild connectivity conditions~\cite{geman1984annealing}.  A \emph{cooling schedule} $T_0,T_1,\dots$  slowly decreases temperature and biases the sampling progressively towards low-energy configurations. In the limit of infinitely slow cooling (e.g.\ logarithmic schedules), SA converges to the global optimum in probability, but practical implementations use finite schedules with limited sweeps over the bitstring~\cite{kirkpatrick1983optimization}.  For our benchmark we implement SA directly on the QUBO, using single-bit neighborhoods, geometric cooling schedules, and multiple random restarts.  SA related metrics are collected in Figs.  \ref{fig:metrics_gallery} and \ref{fig:eta_threeplots}. 

\subsection{Memory-based deterministic Tabu search}

Tabu search is a deterministic local-search meta-heuristic that augments greedy descent with adaptive memory in order to escape local minima~\cite{glover1986tabu}.  On the QUBO energy $E(\mathbf x)$ we define a neighborhood $N(\mathbf x)$ (here, Hamming-1 neighbors).  Starting from $\mathbf x^{(0)}$ we iterate on the following steps:
\begin{enumerate}[leftmargin=1.7em]
  \item generate all candidate moves $m \in N(\mathbf x^{(k)})$ and their energy changes $\Delta E(m)$;
  \item discard moves that are \emph{tabu}, i.e.\ whose inverse move is stored in a short-term tabu list $\mathcal T$ of bounded length $L$;
  \item among the remaining moves, select the one with lowest resulting energy, even if $\Delta E>0$;
  \item update $\mathcal T$ by inserting the chosen move (or its inverse) and possibly removing the oldest entry.
\end{enumerate}
An \emph{aspiration criterion} allows tabu moves that yield a new global best energy to override tabu status.  The tabu list prevents short cycles and encourages exploration of new regions of the energy landscape, while the greedy choice enforces intensification around promising basins~\cite{glover1986tabu}. Within our benchmark we run tabu search directly on the QUBO with fixed tabu tenure $L$ and a cap on the number of iterations per run.  Both Tabu and SA are implemented with D-Wave SDK\cite{mcgeoch2014adiabatic}.

\subsection{Quantum annealing (QA)}
\label{sec:qa}

Quantum annealing (QA) is the quantum analogue of SA, implemented as a continuous-time evolution under a time-dependent Hamiltonian~\cite{kadowaki1998quantum,mcgeoch2014adiabatic,albash2018adiabatic}.  We first map the QUBO to an Ising model in Pauli-$Z$ variables.  Introducing spin variables $z_i \in \{-1,+1\}$ related to the bits by
\(
x_i = \tfrac{1}{2}(1 - z_i)
\). As worked out in Eq. \eqref{eq:quboe}, the QUBO energy can be rewritten (up to an additive constant) as
\begin{equation}
  E(\mathbf x)
  \;=\;
  \mathrm{const}
  \;+\;
  \sum_{i} h_i z_i
  \;+\;
  \sum_{i<j} J_{ij}\,z_i z_j,
  \label{eq:ising-mapping}
\end{equation}
which defines the problem Hamiltonian
\begin{equation}
  H_{\mathrm{P}}
  \;=\;
  \sum_{i} h_i\,\sigma_i^{z}
  \;+\;
  \sum_{i<j} J_{ij}\,\sigma_i^{z}\sigma_j^{z}.
  \label{eq:problem-hamiltonian}
\end{equation}
Here $\sigma_i^{z}$ is the Pauli-$Z$ operator on qubit $i$.  QA adds a transverse-field driver
\begin{equation}
  H_{\mathrm{D}} \;=\; - \sum_{i} \sigma_i^{x},
  \label{eq:driver-hamiltonian}
\end{equation}
and interpolates between them over annealing time $T_{\mathrm{anneal}}$ via
\begin{equation}
  H(s)
  \;=\;
  A(s)\,H_{\mathrm{D}} + B(s)\,H_{\mathrm{P}},
  \qquad
  s = t/T_{\mathrm{anneal}} \in [0,1],
  \label{eq:annealing-schedule}
\end{equation}
with $A(0)\gg B(0)$ and $A(1)\ll B(1)$.  If the evolution is sufficiently slow and the minimum spectral gap is not exponentially small, the adiabatic theorem predicts that the system ends close to the ground state of $H_{\mathrm{P}}$~\cite{albash2018adiabatic}.

On D-Wave-like hardware, QA is realized as an open-system analogue of this process with fixed chip topology and limited coupling precision~\cite{mcgeoch2014adiabatic}.  Our QUBOs are embedded onto the hardware graph and we sample bitstrings from the final measurement distribution, following the benchmarking philosophy of Refs.~\cite{ronnow2014speedup,vert2021benchmarking}.  Repeated shots yield an empirical $p_{\mathrm{succ}}$ and thus STS via Eq.~\eqref{eq:sts-def}.

\noindent\textbf{Remark on quantum annealing runs.}
For $n=4$, QA (Refer Sec. \ref{digression} for tabu) did not produce any feasible solution within the allotted shot budget.
Similarly, for $n\ge 5$, we observed no feasible solutions from QA samples; this likely reflects the combined effect of
(i) embedding-related chain breaks (which become more prevalent as problem size increases) and
(ii) the limited number of shots available under the account used for these runs.
We therefore postpone a more extensive QA evaluation to future work.

\subsection{Leap Hybrid Solver}

The Leap Hybrid solver is a quantum-classical hybrid workflow provided via D-Wave’s Leap service~\cite{dwaveleap}.  It accepts QUBO (or more general quadratic models) at the user interface and internally orchestrates a decomposition across classical and quantum resources~\cite{mcgeoch2014adiabatic,vert2021benchmarking}.  Conceptually, a large QUBO problem is partitioned into overlapping subproblems that fit onto the Quantum processing unit (QPU).  The procedure involves three core steps:
\begin{enumerate}[leftmargin=1.7em]
  \item \textbf{Decomposition:} partition the variable set into overlapping blocks and define sub-QUBOs;
  \item \textbf{Subproblem optimization:} repeatedly send sub-QUBOs to the QPU (quantum annealing) and/or classical heuristics to propose improved assignments;
  \item \textbf{Reconstruction:} reconcile subproblem solutions into a global assignment $\mathbf x$ using classical coordination logic.
\end{enumerate}
The hybrid engine automatically manages the trade-off between quantum calls and classical refinement subject to a user-specified wall-clock limit and returns a candidate bitstring and energy~\cite{dwaveleap}.  Similarly, we also record $E_{\mathrm{hyb}}$, $p_{\mathrm{succ}}$, STS and TTS in in Figs.  \ref{fig:metrics_gallery} and \ref{fig:eta_threeplots}.

\subsection{QAOA: gate-model quantum optimization}

Similar to quantum annealing, the Quantum Approximate Optimization Algorithm (QAOA) is a gate-model, hybrid quantum-classical algorithm that alternates between unitaries generated by the problem Hamiltonian (Eq. \eqref{eq:quboe}) and a transverse-field mixing Hamiltonian (Eq. \eqref{eq:driver-hamiltonian}). It was first proposed in ~\cite{farhi2014qaoa}.  Using the QUBO-derived problem Hamiltonian, conventionally denoted as $H_C$, and the driver Hamiltonian (conventionally denoted  as $H_{\mathrm{M}}$ ), the algorithm prepares the initial state
\(
\ket{\psi_0} = \ket{+}^{\otimes N}
\)
and defines the unitary operators:
\begin{align}
  U_{\mathrm{C}}(\gamma_\ell)
  &= \exp\bigl(-i\,\gamma_\ell H_{\mathrm{C}}\bigr),
  \label{eq:qaoa-cost-unitary}
  \\
  U_{\mathrm{M}}(\beta_\ell)
  &= \exp\bigl(-i\,\beta_\ell H_{\mathrm{M}}\bigr),
  \qquad
  H_{\mathrm{M}} = \sum_{i} \sigma_i^{x},
  \label{eq:qaoa-mixer-unitary}
\end{align}
where $\ell=1,\dots,p$ is the number of alternating layers defined by the user.  The depth-$p$ QAOA state is then
\begin{equation}
  \ket{\boldsymbol{\gamma},\boldsymbol{\beta}}
  \;=\;
  U_{\mathrm{M}}(\beta_p) U_{\mathrm{C}}(\gamma_p)
  \cdots
  U_{\mathrm{M}}(\beta_1) U_{\mathrm{C}}(\gamma_1)
  \ket{\psi_0},
  \label{eq:qaoa-state}
\end{equation}
with parameter vectors
\(
\boldsymbol{\gamma} = (\gamma_1,\dots,\gamma_p),
\;
\boldsymbol{\beta} = (\beta_1,\dots,\beta_p)
\). A classical outer-loop optimizer updates $(\boldsymbol{\gamma},\boldsymbol{\beta})$ to minimize the objective
\begin{equation}
  F_p(\boldsymbol{\gamma},\boldsymbol{\beta})
  \;:=\;
  \bra{\boldsymbol{\gamma},\boldsymbol{\beta}} H_{\mathrm{C}}
  \ket{\boldsymbol{\gamma},\boldsymbol{\beta}},
  \label{eq:qaoa-cost-function}
\end{equation}
which is estimated from repeated measurements of the quantum circuit~\cite{farhi2014qaoa}.  For each near-optimal parameter set, one runs QAOA in ``sampling mode'' to generate bitstrings which represent the candidate solutions. Several recent studies have significantly expanded and enhanced this basic formalism. \cite{hadfield2019qaoa,zhou2020quantum,harwood2021formulating,onahce}. 

Our first QAOA implementation follows the linear ramp schedules introduced in Ref. \cite{montanezba} to deterministically  fix the hyper parameters $(\gamma_l,\beta_l)$ in Eq. \eqref{eq:qaoa-state} and avoid the classical parameter optimization overhead. Our second implementation follows a targeted design principle described in~\cite{onahce}.

\subsubsection{Linear-Ramp QAOA (LR-QAOA)}

In the standard formalism, the parameter vector
$(\boldsymbol{\gamma},\boldsymbol{\beta})$ of a depth-$p$ QAOA circuit in
Eq.~\eqref{eq:qaoa-state} has $2p$ free angles.  In  Linear-Ramp QAOA (LR-QAOA), the parameter vector is reduced to two \emph{ramp parameters}
$\Delta\gamma,\Delta\beta$ by imposing a linear schedule over the
layers ~\cite{montanezba},
\begin{equation}
  \gamma_\ell
  \;=\;
  \frac{\ell}{p}\,\Delta\gamma,
  \qquad
  \beta_\ell
  \;=\;
  \frac{p{+}1-\ell}{p}\,\Delta\beta,
  \qquad
  \ell = 1,\dots,p.
  \label{eq:lr-qaoa-schedule}
\end{equation}
The cost angle is ramped up linearly while the mixer angle is
ramped down linearly, in direct analogy with a discretized annealing
schedule.  The depth-$p$ LR-QAOA state is still given by
Eq.~\eqref{eq:qaoa-state}, but is now completely specified by the
two-dimensional vector
$\boldsymbol{\theta}_{\mathrm{LR}}=(\Delta\gamma,\Delta\beta)$,
independent of $p$.

\begin{figure}[htbp]
  \centering
  \begin{tikzpicture}
    \begin{axis}[
      width=0.7\linewidth,
      height=5cm,
      xlabel={layer index $\ell$},
      ylabel={angle value},
      xmin=1, xmax=6,
      ymin=0, ymax=1.2,
      xtick={1,2,3,4,5,6},  
      ytick=\empty,
      axis lines=left,
      grid=both,
      grid style={densely dotted},
      legend style={at={(0.5,1.02)}, anchor=south, legend columns=2},
      every axis plot/.append style={very thick}
    ]

      \addplot[solid] coordinates {
        (1,0.2)
        (2,0.4)
        (3,0.6)
        (4,0.8)
        (5,1.0)
        (6,1.2)
      };
      \addlegendentry{$\gamma_{\ell}$}

      \addplot[dashed] coordinates {
        (1,0.8)
        (2,0.7)
        (3,0.6)
        (4,0.5)
        (5,0.4)
        (6,0.3)
      };
      \addlegendentry{$\beta_{\ell}$}

    \end{axis}
  \end{tikzpicture}
  \caption{Schematic linear--ramp QAOA schedule. The cost angles
  $\gamma_{\ell}$ increase linearly with the layer index $\ell$ while the
  mixer angles $\beta_{\ell}$ decrease linearly without instance--specific angle optimization.}
  \label{fig:lr-qaoa-schedule}
\end{figure}
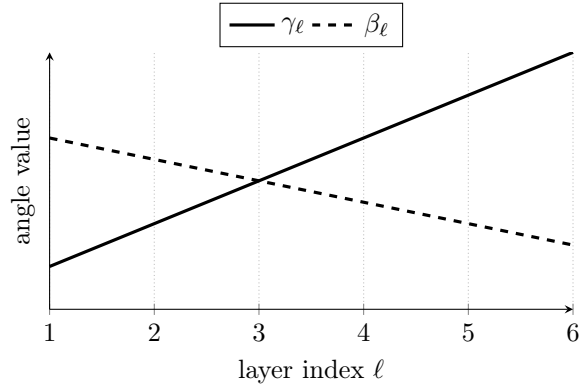

with a discretised adiabatic path.

\begin{figure}[htbp]
  \centering
  \includegraphics[width=0.6\linewidth]{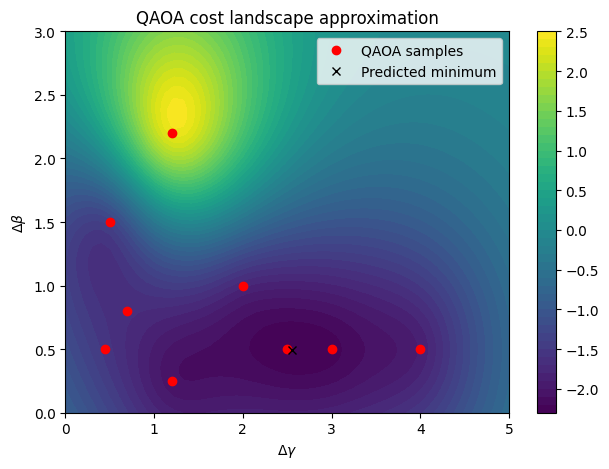}
  \caption{QAOA cost landscape approximation for the linear--ramp schedule.
  The contour plot shows the Gaussian–process surrogate of the depth-$p$
  objective $F_p(\Delta\gamma,\Delta\beta)$ over the two ramp parameters.
  Red markers indicate the QAOA evaluations used to train the surrogate,
  while the black cross marks the predicted minimum $(\Delta\gamma^\star,
  \Delta\beta^\star)$ that defines the final linear schedule.}
  \label{fig:lr-qaoa-landscape}
\end{figure}

In our implementation, for a fixed depth $p$, we estimate the expected energy
\(
  F_p(\Delta\gamma,\Delta\beta)
\) (Eq. 
\eqref{eq:qaoa-cost-function})
by sampling the circuit generated by the schedule
(Eq. \eqref{eq:lr-qaoa-schedule}).  A Gaussian-process Bayesian optimizer
over $(\Delta\gamma,\Delta\beta)$ is then used to locate a good
minimum of $F_p$ on a coarse grid.  The resulting $(\Delta\gamma^\star,\Delta\beta^\star)$ are
kept fixed for all subsequent sampling runs on that instance.  This
reduces the classical optimization overhead from a $2p$-dimensional
search to a two-dimensional calibration problem, while preserving a hardware-efficient layered circuit compatible with gate-based
superconducting devices.

\begin{figure*}[t]
  \centering
  \includegraphics[width=\textwidth]{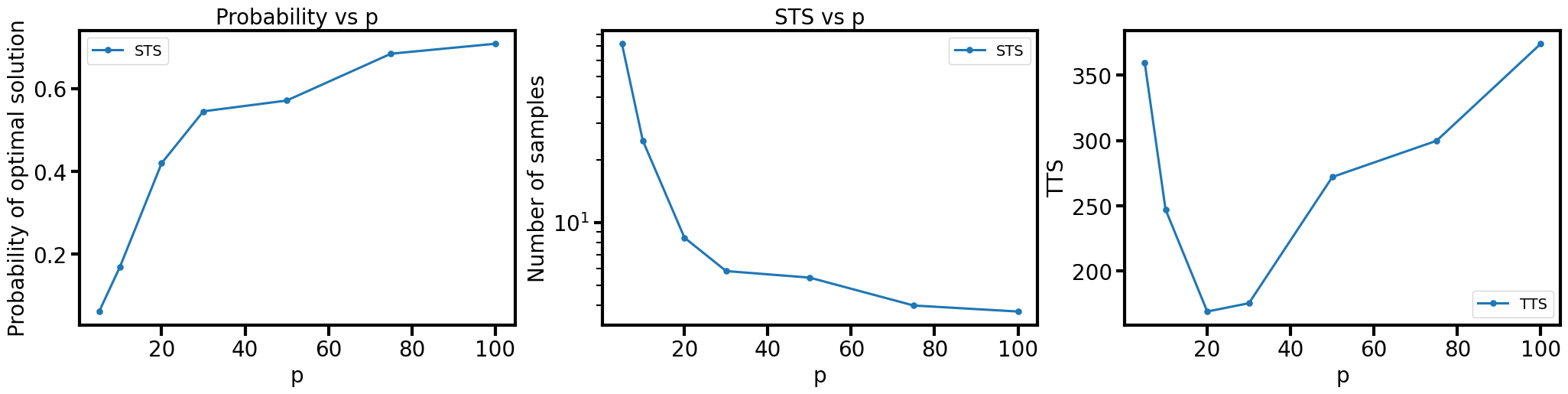}
  \caption[Linear–ramp QAOA performance vs depth]{%
    Linear–ramp QAOA performance as a function of circuit depth $p$
    for a $4\times 4$ instance (four surfers matched to four breakers).
    Left: single–shot success probability $p_{\mathrm{succ}}$ increases roughly
    monotonically with $p$, as expected from a more adiabatic schedule.
    Middle: samples–to–solution (STS) on a log scale drops by more than an order
    of magnitude before saturating, indicating improved sample efficiency at
    larger depth. Right: time–to–solution (TTS) shows a broad plateau, suggesting
    that beyond moderate depths additional layers do not pay off in wall–clock
    time even though $p_{\mathrm{succ}}$ continues to improve.}
  \label{fig:lrqaoa-depth-sweep}
\end{figure*}

LR-QAOA reduces much of the heuristic “guesswork’’ in variational quantum optimization and yields a hardware–friendly protocol whose complexity is controlled almost entirely by the achievable depth and coherence time. As devices support larger $p$, the same scheduling rule extends seamlessly to deeper circuits, making LR–QAOA an attractive baseline for large–scale, low–tuning quantum optimization. At the same time, because the schedule is expressed in a low–dimensional family of ramps, it remains compatible with warm–starting and parameter–transfer strategies \cite{montanezbarrera2024transfer}. Precomputed QAOA parameters (e.g.\ from small instances, mean–field limits, or from single layer QAOA which is classically solvable \cite{onah2025probe,onah2025params}) can be distilled into effective ramp slopes and offsets, allowing LR–QAOA to inherit benefits from the broader theory of QAOA parameter transfer while keeping the on–device optimization overhead minimal.

\subsubsection{Constraint Enhanced QAOA (CE-QAOA)}
\label{sec:ce-qaoa}
Our second QAOA variant is the Constraint Enhanced QAOA introduced in \cite{onahce}. It follows the general problem-algorithm co-design principles to build the quantum algorithm around the problem structure. An adaptation to the two-vehicle matching problem is as follows: let $S$ be the number of surfers and $B$ be the number of breakers. We group qubits into blocks such that each block has $S$ qubits and we have $B$ blocks in total. The total number of qubits to encode the problem  remains the same ($S\cdot B$) as in the previous implementation \cite{onah2025}. However, in this case, a dedicated mixer and initial state is defined over each block such that each block provides a single assignment possibility. The overall synchronization to maintain feasibility is enforced by the breaker uniqueness constraint in the diagonal Hamiltonian. Precisely, on each block we define,

\begin{equation}
\begin{aligned}
      \widetilde H_{XY}:=\frac{1}{S-1}\sum_{a<b}\bigl(X_aX_b+Y_aY_b\bigr),\qquad
      U_M(\beta)=\bigotimes_{b=1}^B e^{-i\beta \widetilde H^{(b)}_{XY}},
      \\
      \text{and} \ket{s_0}=\ket{s_{\mathrm{blk}}}^{\otimes B} \text{with} 
      \ket{s_{\mathrm{blk}}}=\tfrac1{\sqrt S}\sum_{j=1}^S\ket{e_j}.
\end{aligned}
\label{eq:hamiltonian_ceqaoa}
\end{equation}

\begin{figure}[ht]
  \centering
  \resizebox{\textwidth}{!}{%
  \begin{quantikz}[row sep=0.12cm, column sep=0.1cm]
  \lstick[wires=4]{$\text{Block }0:\ket{0}^{\otimes S}$}
      & \gate[wires=4,style={rounded corners,fill=blue!8}]
          {\texttt{OneHotBlock}}
      & \qw
      & \gate[wires=16,style={rounded corners,fill=green!16}]
          {U_C(\gamma_1)}
      & \gate[wires=4,style={rounded corners,fill=orange!20}]
          {U_M^{(0)}(\beta_1)}
      & \qw
      & \gate[wires=16,style={rounded corners,fill=green!16}]
          {U_C(\gamma_2)}
      & \gate[wires=4,style={rounded corners,fill=orange!20}]
          {U_M^{(0)}(\beta_2)}
      & \qw
      & \gate[wires=16,style={rounded corners,fill=green!16}]
          {U_C(\gamma_3)}
      & \gate[wires=4,style={rounded corners,fill=orange!20}]
          {U_M^{(0)}(\beta_3)}
      & \qw
      & \meter{} \\
  & & \qw
    & \qw & \qw & \qw
    & \qw & \qw & \qw
    & \qw & \qw & \qw
    & \meter{} \\
  & & \qw
    & \qw & \qw & \qw
    & \qw & \qw & \qw
    & \qw & \qw & \qw
    & \meter{} \\
  & & \qw
    & \qw & \qw & \qw
    & \qw & \qw & \qw
    & \qw & \qw & \qw
    & \meter{} \\
  \lstick[wires=4]{$\text{Block }1:\ket{0}^{\otimes S}$}
      & \gate[wires=4,style={rounded corners,fill=blue!8}]
          {\texttt{OneHotBlock}}
      & \qw
      & \qw
      & \gate[wires=4,style={rounded corners,fill=orange!20}]
          {U_M^{(1)}(\beta_1)}
      & \qw
      & \qw
      & \gate[wires=4,style={rounded corners,fill=orange!20}]
          {U_M^{(1)}(\beta_2)}
      & \qw
      & \qw
      & \gate[wires=4,style={rounded corners,fill=orange!20}]
          {U_M^{(1)}(\beta_3)}
      & \qw
      & \meter{} \\
  & & \qw
    & \qw & \qw & \qw
    & \qw & \qw & \qw
    & \qw & \qw & \qw
    & \meter{} \\
  & & \qw
    & \qw & \qw & \qw
    & \qw & \qw & \qw
    & \qw & \qw & \qw
    & \meter{} \\
  & & \qw
    & \qw & \qw & \qw
    & \qw & \qw & \qw
    & \qw & \qw & \qw
    & \meter{} \\
  \lstick[wires=4]{$\text{Block }2:\ket{0}^{\otimes S}$}
      & \gate[wires=4,style={rounded corners,fill=blue!8}]
          {\texttt{OneHotBlock}}
      & \qw
      & \qw
      & \gate[wires=4,style={rounded corners,fill=orange!20}]
          {U_M^{(2)}(\beta_1)}
      & \qw
      & \qw
      & \gate[wires=4,style={rounded corners,fill=orange!20}]
          {U_M^{(2)}(\beta_2)}
      & \qw
      & \qw
      & \gate[wires=4,style={rounded corners,fill=orange!20}]
          {U_M^{(2)}(\beta_3)}
      & \qw
      & \meter{} \\
  & & \qw
    & \qw & \qw & \qw
    & \qw & \qw & \qw
    & \qw & \qw & \qw
    & \meter{} \\
  & & \qw
    & \qw & \qw & \qw
    & \qw & \qw & \qw
    & \qw & \qw & \qw
    & \meter{} \\
  & & \qw
    & \qw & \qw & \qw
    & \qw & \qw & \qw
    & \qw & \qw & \qw
    & \meter{} \\
  \lstick[wires=4]{$\text{Block }3:\ket{0}^{\otimes S}$}
      & \gate[wires=4,style={rounded corners,fill=blue!8}]
          {\texttt{OneHotBlock}}
      & \qw
      & \qw
      & \gate[wires=4,style={rounded corners,fill=orange!20}]
          {U_M^{(3)}(\beta_1)}
      & \qw
      & \qw
      & \gate[wires=4,style={rounded corners,fill=orange!20}]
          {U_M^{(3)}(\beta_2)}
      & \qw
      & \qw
      & \gate[wires=4,style={rounded corners,fill=orange!20}]
          {U_M^{(3)}(\beta_3)}
      & \qw
      & \meter{} \\
  & & \qw
    & \qw & \qw & \qw
    & \qw & \qw & \qw
    & \qw & \qw & \qw
    & \meter{} \\
  & & \qw
    & \qw & \qw & \qw
    & \qw & \qw & \qw
    & \qw & \qw & \qw
    & \meter{} \\
  & & \qw
    & \qw & \qw & \qw
    & \qw & \qw & \qw
    & \qw & \qw & \qw
    & \meter{} \\
  \end{quantikz}
  }
  \caption{Depth-\(p=3\) CE--QAOA for \(B=4\) blocks of \(S=4\) qubits
  (total \(SB=16\) qubits). Each block corresponds to one breaker and
  encodes a one-hot superposition \(\ket{s_{\mathrm{blk}}}\) over the
  \(S\) surfer indices prepared by \texttt{OneHotBlock}, so that the
  global initial state is
  \(\ket{s_0}=\ket{s_{\mathrm{blk}}}^{\otimes B}\).
  Each layer applies a global cost unitary
  \(U_C(\gamma_\ell)=e^{-i\gamma_\ell H_C}\) over all \(SB\) wires,
  followed by parallel block-local XY mixers
  \(U_M^{(b)}(\beta_\ell)=\exp[-i\beta_\ell \tilde H_{XY}^{(b)}]\) on
  each block \(b=0,1,2,3\), with
  \(\tilde H_{XY}=\tfrac{1}{S-1}\sum_{a<b}(X_aX_b+Y_aY_b)\).}
  \label{fig:full-Blockqaoa}
\end{figure}
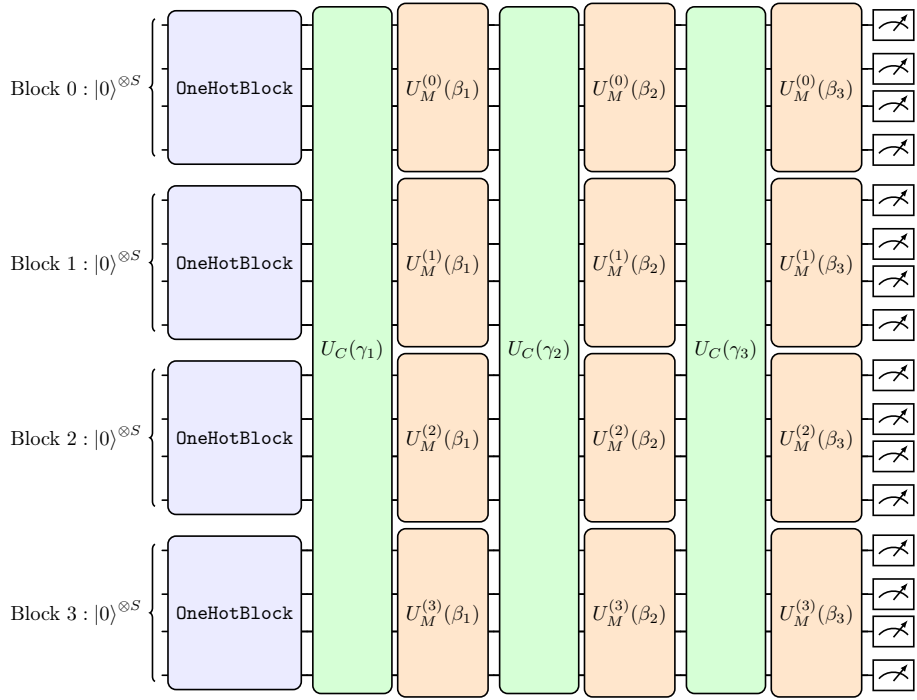

\begin{figure}[htbp]
  \centering
  \begin{subfigure}[t]{0.32\textwidth}
    \centering
    \includegraphics[width=\linewidth]{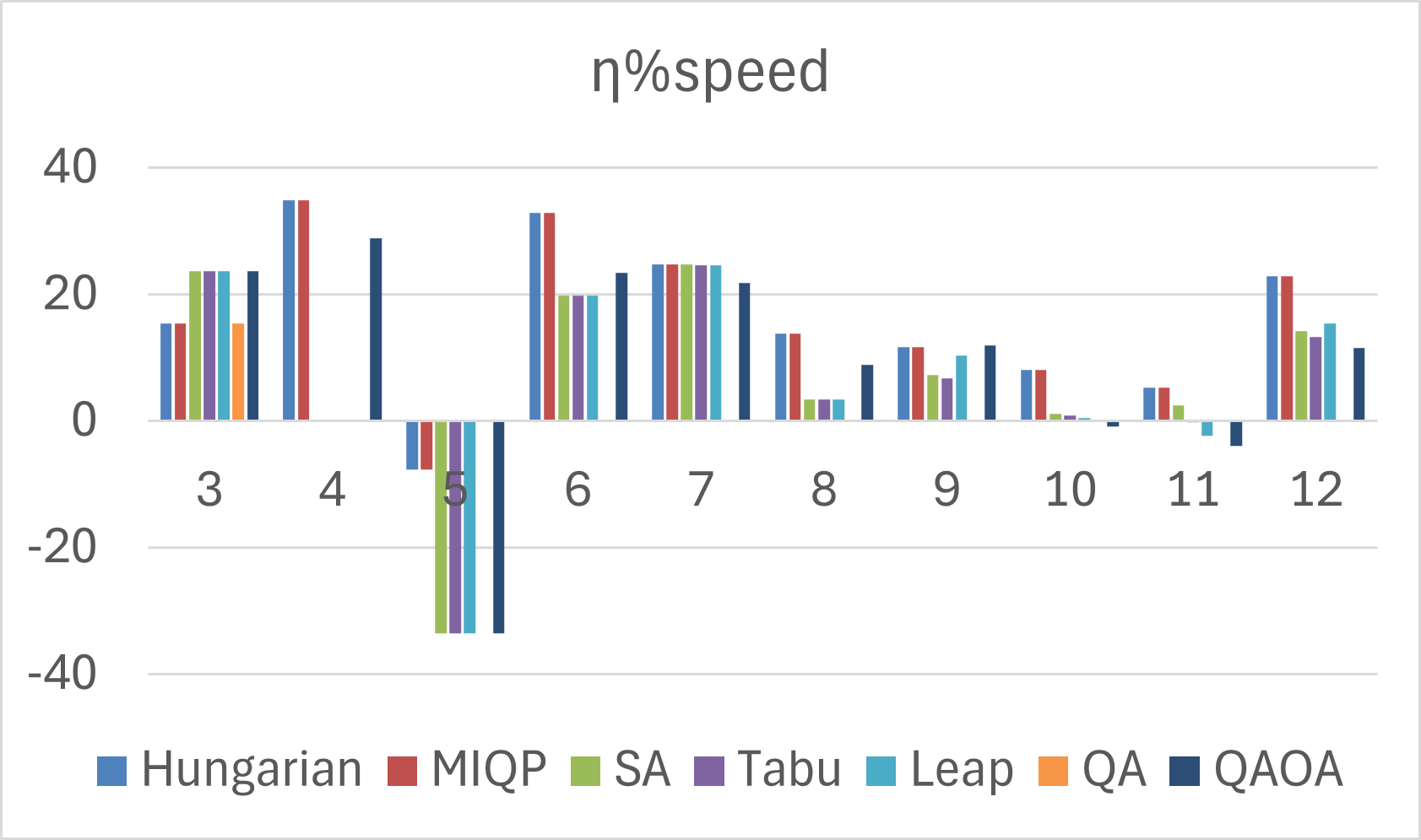}
    \caption{$\eta_{\mathrm{speed}}$}
    \label{fig:eta_speed}
  \end{subfigure}\hfill
  \begin{subfigure}[t]{0.32\textwidth}
    \centering
    \includegraphics[width=\linewidth]{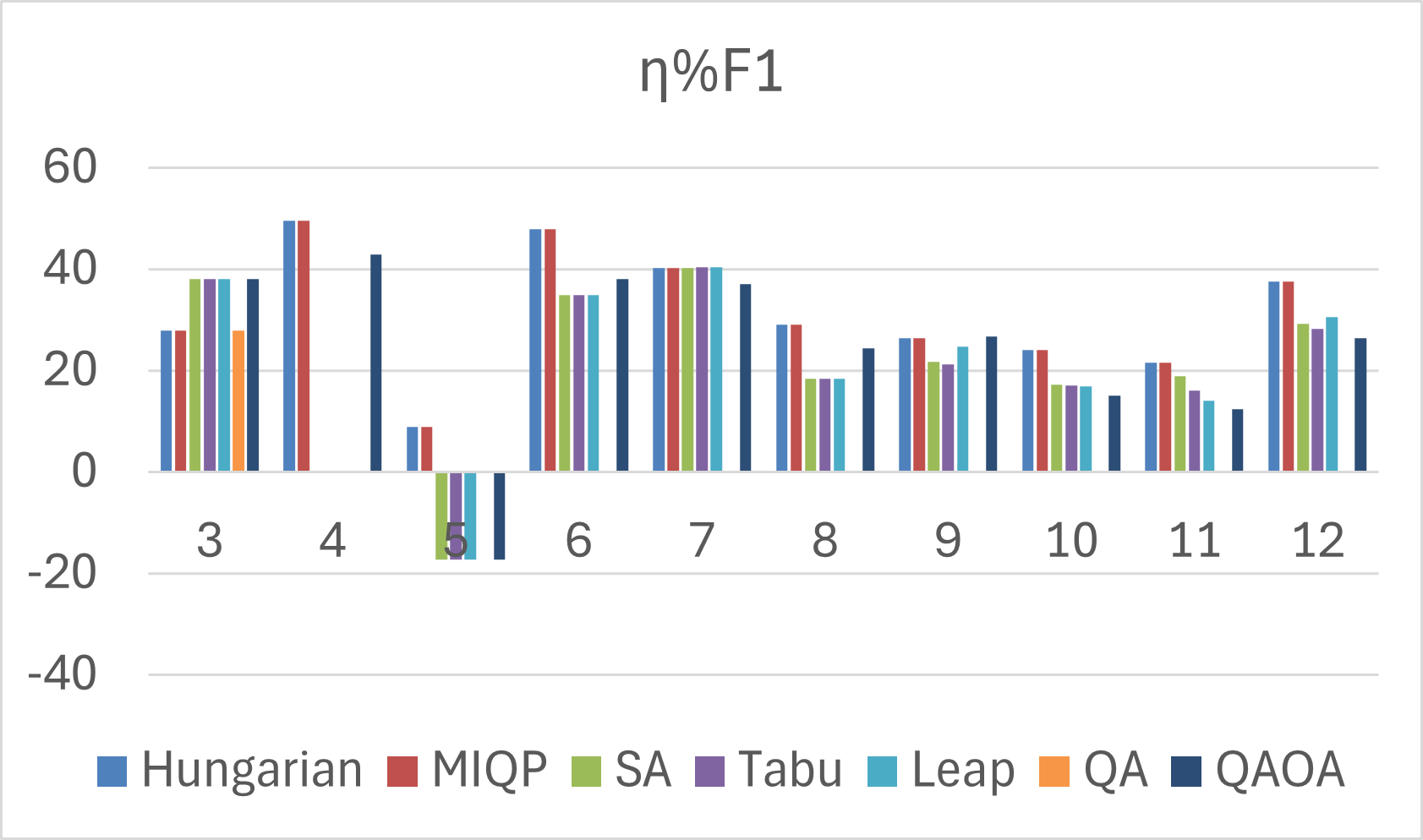}
    \caption{$\eta_{F_1}$}
    \label{fig:eta_F1}
  \end{subfigure}\hfill
  \begin{subfigure}[t]{0.32\textwidth}
    \centering
    \includegraphics[width=\linewidth]{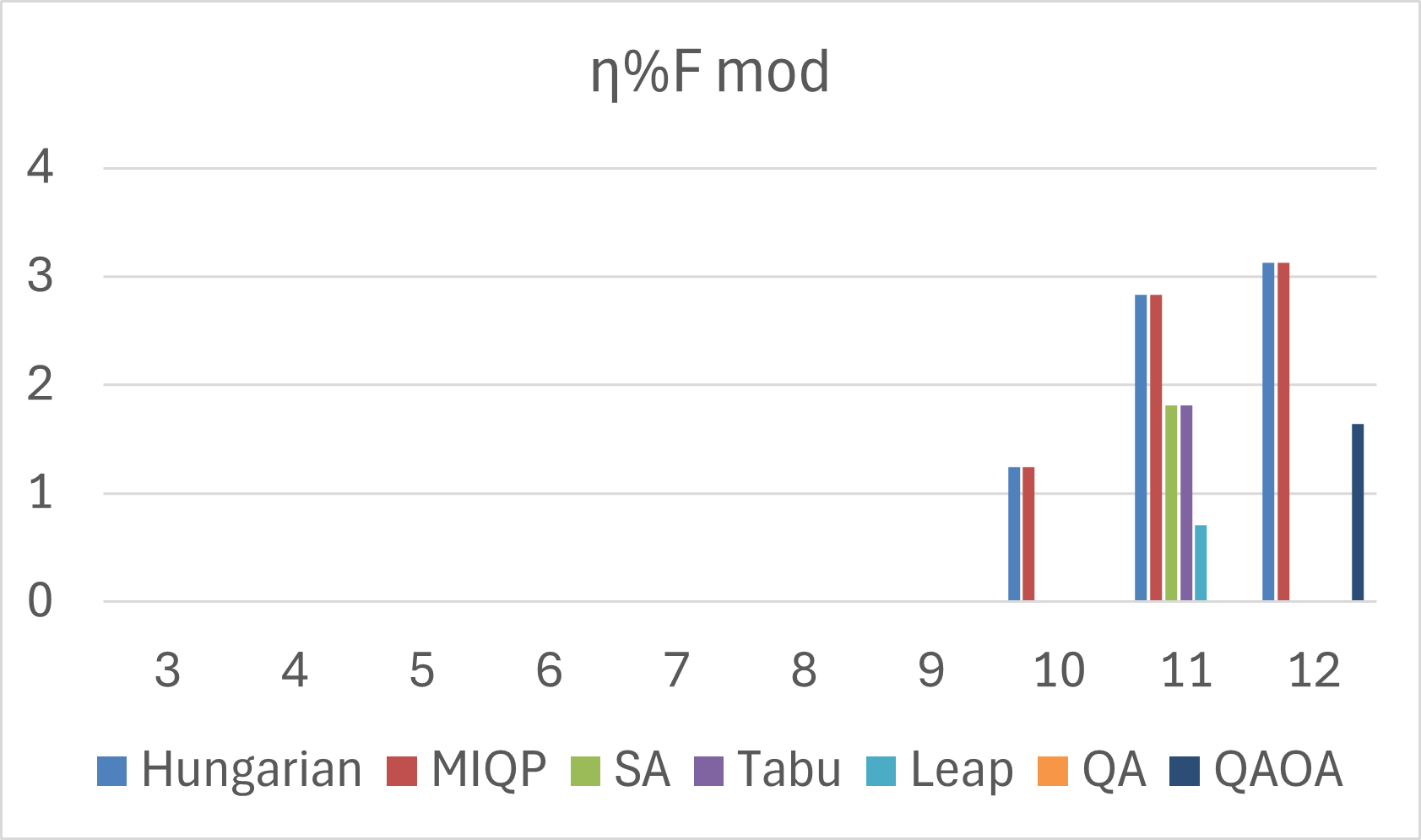}
    \caption{$\eta_{F_1^{\mathrm{mod}}}$}
    \label{fig:eta_F1mod}
  \end{subfigure}

  \caption{Comparison of percentage savings metrics relative to $F_{\mathrm{ref}}$.  Negative values in $\eta_{\mathrm{speed}}$ and $\eta_{F_1}$ can occur when an assignment forces a surfer to drive faster than their preferred speed ($V_b>v_s$), so the quadratic speed scaling increases energy enough to outweigh slipstream gains and yields worse-than-solo performance.
  The post-optimization veto used in $F_1^{\mathrm{mod}}$ removes this pathology by letting surfers reject such undesirable pairings and revert to solo driving at $v_s$, so $\eta_{F_1^{\mathrm{mod}}}$ reflects realizable savings under human acceptance. See the underlying tables in the Supplementary Material in App. \ref{app:app}.}
  \label{fig:eta_threeplots}
\end{figure}

The problem Hamiltonian is the same as in Eq. \eqref{eq:quboe} and the samples generated from the quantum measurements in the computational basis are passed through a classical checker that returns the best feasible solution. The algorithm has feasibility guarantee built in \cite{onahce,onah2025empiricaldata}. The classical parameter optimization is replaced by a finite grid search whose resolution is defined by $B$. A running example of $S=B=4$ circuit is exhibited in Fig \ref{fig:full-Blockqaoa}. Figs.~\ref{fig:metrics_gallery} and \ref{fig:eta_threeplots} summarizes the empirical performance of a single layer of CE--QAOA across all instances relative to other samplers studied in this work. 

\paragraph{Note:} We simulated CE-QAOA circuits for instance sizes ranging from $n=3$ vehicle pairs ($N=n^2=9$ qubits) up to $n=12$ vehicle pairs ($N=144$ qubits). To keep the state-vector simulation tractable at these widths, we used the Aer \texttt{matrix\_product\_state} backend with MPS truncation enabled (\texttt{matrix\_product\_state\_max\_bond\_dimension}=128 and truncation/validation thresholds of $10^{-3}$). For each parameter setting, the circuit was sampled to produce bitstrings, which were then post-processed classically. Samples were filtered by the permutation-matrix feasibility constraints and the feasible subset was evaluated under the same objective/Ising energy to report the best feasible sample returned by the quantum run.

\subsection{A Digression on local trapping.}\label{digression}
While the Tabu/SA/Leap solvers produced feasible samples for most problem sizes, we observed a pathological trapping behavior for the instance with \(n=4\). 
In this case, a run with \(N_{\mathrm{reads}}=320\) returned only two unique bitstrings and \emph{no} feasible samples (i.e., \(\hat p_{\mathrm{feas}}=0\)), so the empirical success probability was necessarily \(\hat p_{\mathrm{succ}}=0\) under our definition (success requires feasibility and optimality within tolerance). 
This indicates that the Tabu dynamics became confined to a very small region of the unconstrained hypercube and failed to reach the permutation manifold (row/column one-hot constraints) at all. 
By contrast, for \(n\in\{3,5,6,\ldots\}\) the same implementation, typically generated feasible samples in every read, often with low diversity (e.g., a single unique feasible bitstring repeated), suggesting that once a feasible basin is entered the heuristic can remain trapped there as well.

We further found that increasing the penalty scaling parameter \(\lambda_3\) can force the Tabu heuristic to produce feasible assignments for \(n=4\), but at the cost of yielding systematically suboptimal objective values, consistent with an overly strong penalty dominating the effective energy landscape. 
Notably, the theoretically-motivated scaling used throughout this work produced stable behavior across the remaining instances, which suggests that the \(n=4\) failure mode is more\, likely attributable to heuristic initialization and/or Tabu hyperparameters (e.g., initial state, tenure, or neighborhood/aspiration settings) than to a fundamental issue with the formulation. 
Since Tabu Search is used here purely as a reference heuristic and comprehensive hyperparameter optimization is not the focus of this paper, we leave a systematic parameter and restart study (including multi-start initialization strategies) for future work.

\section{Evaluating Practical Benefits}
\label{sec:cost-to-impact}

Optimizing Windbreaking-as-a-Service (WaaS) is meaningful only if the
binary decisions inside the QUBO translate into transparent physical
benefits like \(\mathrm{CO_2}\) emission reduction, EV-range extension, and charger relief. For a windsurfer of class \(c_s\) drafting behind a breaker of class,
\(C_b\) the slip-stream efficiency is modeled as the linear map
\cite{alam2015review,onah2025}
\begin{equation}\label{eq:f}
  f(d)\;=\;\frac{d+4}{24},
  \qquad
  d:=C_b-c_s\in\{-4,\dots,4\},
\end{equation}

so that \(f(-4)=0\) (tiny breaker, large surfer) and \(f(4)=\tfrac13\)
(best pairing: large breaker, small surfer).


The surfer classes are defined such that an individual vehicle's aerodynamic drag coefficient is (roughly) proportional to the class $c_s$ it belongs to. Therefore, if a surfer $s$ is traversing a certain motorway edge 
at the breaker’s speed \(V_b\), the associated aerodynamic \emph{energy  per unit distance} is
\begin{equation}\label{eq:E_edge}
  E_{s,b} = c_s\,V_b^{2}\,\bigl[1-f(C_b-c_s)\bigr].
\end{equation}



To minimize the aerodynamic drag energy and, by extension, \(\mathrm{CO_2}\) emissions
or battery discharge, we track \emph{four} cost functions outlined below (also summarized in Table \ref{tab:costs}).

\paragraph{1 — Solo driving at preferred speed.}
\label{sec:Fref}

The \emph{reference cost}

\begin{equation}
  F_{\mathrm{ref}}
  =\sum_{s=1}^{S} c_s\,v_s^{2}
  \label{eq:Fref}
\end{equation}

assumes every windsurfer drives alone at their own cruise speed \(v_s\).
All subsequent energies are reported \emph{relative} to \(F_\text{ref}\) 

\paragraph{2 — Velocity-corrected reference.}
\label{sec:FrefVel}

 The problem setup is such that the surfers are forced to follow the breakers at their speed \(V_b\neq v_s\).  The hypothetical cost of solo vehicle driving at corrected velocity \(V_b\) is

\begin{equation}
  F_{\mathrm{ref}}^{\mathrm{vel}}
  =\sum_{s=1}^{S} c_s\,V_b^{2} .
  \label{eq:FrefVel}
\end{equation}

Comparing \(F_{\mathrm{ref}}^{\mathrm{vel}}\) with \(F_{\mathrm{ref}}\)
\emph{isolates} the energetic penalty (or gain) is due purely to imposed
speed changes and is crucial when the optimizer produces very slow or
very fast convoys.

\paragraph{3 — Objective part of the cost minimized by the optimizer}
\label{sec:F1}

The matching problem minimizes

\begin{equation}
  F_1
  =\sum_{s=1}^{S}
    c_s\,\,V_b^{2}
    \bigl[1-f(C_b-c_s)\bigr],
  \label{eq:F1_main}
\end{equation}

i.e.\ aerodynamic energy with the breaker speed and the slipstream
efficiency from Eq.~\eqref{eq:f} with every computed pairing accepted.

\paragraph{4 — Post Optimization: modified objective}
\label{sec:F1mod}

After optimization we re-evaluate the schedule: any surfer whose timing
or velocity constraints are still violated simply \emph{opts out} and
keeps driving alone at \(v_s\).  Splitting the index set into
“assigned’’ surfers \(A\) and “unassigned’’ surfers \(U\) gives

\begin{equation}
  F_1^{\mathrm{mod}}
  =\sum_{s\in A} c_s V_b^{2}\bigl[1-f(C_b-c_s)\bigr]
  +\sum_{s\in U} c_s v_s^{2}.
  \label{eq:F1mod}
\end{equation}

The gap \(F_1^{\mathrm{mod}}-F_{\mathrm{ref}}\) therefore measures the
\emph{realizable} drag saving once human acceptance is considered.

\begin{table}[H]
\centering
\caption{Hierarchy of cost functions.}
\label{tab:costs}
\begin{tabularx}{\textwidth}{c X c}
\toprule
\textbf{Name} & \textbf{Informal meaning} & Eq.\# \\\midrule
\(F_\text{ref}\) 
& Every surfer drives \emph{solo} at their own preferred speed \(v_s\); no windbreaking. & \eqref{eq:Fref} \\
\(F_\text{ref}^{\text{vel}}\) 
& Still solo, but forced to adopt the \emph{speed profile implied by the optimizer} (typically the leader’s speed).  Isolates the pure ‘‘speed-change’’ penalty or gain. & \eqref{eq:FrefVel} \\
\(F_1\) 
& The optimizer’s aerodynamic energy using pairing rule and breaker speed, but \emph{ignoring unmet preferences}.  This is the \textbf{objective actually minimized} in the QUBO/MIQP.\\  

\(F_1^{\text{mod}}\) 

& Post-hoc correction: surfers whose constraints are violated abandon the match and revert to solo driving at \(v_s\).  Gives a realistic \emph{ex-post} energy once humans are allowed to veto bad pairings. & \eqref{eq:F1mod}\\ \bottomrule
\end{tabularx}
\end{table}

The percentage energy change relative to the preferred-speed baseline
$F_{\mathrm{ref}}$ as
\begin{align}
\eta_{\mathrm{speed}}
&:= 100\,\frac{F_{\mathrm{ref}}-F_{\mathrm{ref}}^{\mathrm{vel}}}{F_{\mathrm{ref}}},
\label{eq:eta_speed}\\[2pt]
\eta_{F_1}
&:= 100\,\frac{F_{\mathrm{ref}}-F_{1}}{F_{\mathrm{ref}}},
\label{eq:eta_F1}\\[2pt]
\eta_{F_1^{\mathrm{mod}}}
&:= 100\,\frac{F_{\mathrm{ref}}-F_{1}^{\mathrm{mod}}}{F_{\mathrm{ref}}}.
\label{eq:eta_F1mod}
\end{align}
A positive $\eta$ indicates an energy saving relative to $F_{\mathrm{ref}}$,
while a negative value indicates an energy increase.

\section{Conclusion}
\label{sec:conclusion}

We studied the two-vehicle matching subproblem behind Windbreaking-as-a-Service (WaaS) and cast it into a linear assignment formulation with an equivalent QUBO/Ising representation. This allowed us to benchmark a heterogeneous ``solver zoo'' including exact classical baselines (Hungarian, MIQP), classical meta-heuristics (SA, Tabu), and quantum / quantum-inspired approaches (quantum annealing, Leap Hybrid, gate-model QAOA). The main practical takeaway is that the QUBO acts as a \emph{lingua franca} that lets fundamentally different backends attack the same high-order landscape and makes solver comparisons meaningful via shared solver agnostic metrics. 

Across the tested instances, the exact baselines provide ground-truth labels and confirm correctness of the encodings. Among heuristic methods, we observed that feasibility is not automatic in the unconstrained hypercube and depends sensitively on penalty scaling and solver dynamics. This motivates constraint-aware quantum design like CE--QAOA which enforces row-wise one-hot structure by construction and shifts the burden to a smaller set of remaining constraints plus a classical feasibility checker. Even at shallow depth ($p=1$), this structured approach is observed to yield non-trivial success probabilities, illustrating why problem-aware encodings can matter more than simply increasing circuit depth in generic QAOA. On the other hand, LR--QAOA mitigates the parameter--tuning bottleneck of quantum heuristics by replacing the unstable $2p$-angle search with a fixed monotone schedule, making depth $p$ the main control knob. Figure~\ref{fig:lrqaoa-depth-sweep} shows that this yields a well-behaved scaling where $p_{\mathrm{succ}}$ increases with $p$ and STS falls dramatically, so deeper circuits translate directly into higher hit-rate without the confounder of optimizer convergence. The early TTS plateau then identifies a practical depth window where added layers still improve success probability but no longer buy wall-clock speedups, separating algorithmic gains from hardware overhead.

\section*{Data Availability} All data and Python implementation are available at: \url{https://doi.org/10.5281/zenodo.17768136} \cite{onah2025data}.

\section*{Conflict of Interests}
All authors declare no competing interests.

\section*{Author contributions}
All authors contributed equally. In addition, Carsten Othmer and Kristel Michielsen provided guidance during the project. All the authors reviewed the paper.

\section*{Funding}
This research received no external funding.


\textbf{Correspondence and requests for materials} should be addressed to Chinonso Onah\ (\texttt{chinonso.calistus.onah@volkswagen.de}).




\bibliographystyle{splncs04}
\bibliography{references}

\begin{thebibliography}{10}
\providecommand{\url}[1]{\texttt{#1}}
\providecommand{\urlprefix}{URL }
\providecommand{\doi}[1]{https://doi.org/#1}

\bibitem{alam2015review}
Alam, A., Besselink, B., Turri, V., Mårtensson, J., Johansson, K.~\, H.: Heavy-duty vehicle platooning for sustainable freight transportation: A review. IEEE Control Systems Magazine  \textbf{35}(6),  34--56 (2015)

\bibitem{albash2018adiabatic}
Albash, T., Lidar, D.~\, A.: Adiabatic quantum computation. Reviews of Modern Physics  \textbf{90}(1),  015002 (2018)

\bibitem{farhi2014qaoa}
Farhi, E., Goldstone, J., Gutmann, S.: A quantum approximate optimization algorithm (2014), [quant-ph]

\bibitem{geman1984annealing}
Geman, S., Geman, D.: Stochastic relaxation, gibbs distributions, and the {B}ayesian restoration of images. IEEE Transactions on Pattern Analysis and Machine Intelligence  \textbf{PAMI-6},  721--741 (1984), (6)

\bibitem{glover1986tabu}
Glover, F.: Future paths for integer programming and links to artificial intelligence. Computers \& Operations Research  \textbf{13}(5),  533--549 (1986)

\bibitem{hadfield2019qaoa}
Hadfield, S., Wang, Z., O'Gorman, B., et~al.: From the quantum approximate optimization algorithm to a quantum alternating operator ansatz. Algorithms  \textbf{12}(2), ~34 (2019)

\bibitem{harwood2021formulating}
Harwood, S., Gambella, C., Trenev, D., Simonetto, A., Bernal, D., Greenberg, D.: Formulating and solving routing problems on quantum computers. IEEE transactions on quantum engineering  \textbf{2},  1--17 (2021)

\bibitem{dwaveleap}
Inc, D.W.S.: Leap service's hybrid solvers documentation. (accessed 2025), https://docs.dwavequantum.com

\bibitem{kadowaki1998quantum}
Kadowaki, T., Nishimori, H.: Quantum annealing in the transverse ising model. Physical Review E  \textbf{58}(5),  5355--5363 (1998)

\bibitem{kirkpatrick1983optimization}
Kirkpatrick, S., Gelatt, C.~\, D., Vecchi, M.~\, P.: Optimization by simulated annealing. Science  \textbf{220}(4598),  671--680 (1983)

\bibitem{kuhn1955hungarian}
Kuhn, H.~\, W.: The hungarian method for the assignment problem. Naval Research Logistics Quarterly  \textbf{2}(1-2),  83--97 (1955)

\bibitem{Lucas_2014}
Lucas, A.: Ising formulations of many np problems. Frontiers in physics  \textbf{2}, ~5 (2014)

\bibitem{mcgeoch2014adiabatic}
McGeoch, C.~\, C.: Adiabatic Quantum Computation and Quantum Annealing: Theory and Practice. Synthesis Lectures on Quantum Computing. Morgan \& Claypool, San Rafael (2014)

\bibitem{montanezba}
Montanez-Barrera, J., Michielsen, K.: Towards a universal qaoa protocol: Evidence of a scaling advantage in solving some combinatorial optimization problems. arxiv 2024. arXiv preprint arXiv:2405.09169  (2024)

\bibitem{montanezbarrera2024transfer}
Montañez-Barrera, J.A., Willsch, D., Michielsen, K.: Transfer learning of optimal qaoa parameters in combinatorial optimization (2024), arXiv preprint

\bibitem{onah2025probe}
Onah, C., Michielsen, K.: Scalable hardware maturity probe for quantum accelerators via harmonic analysis of qaoa (2025), [quant-ph]

\bibitem{onah2025params}
Onah, C.: Single-layer qaoa p = 1 global-optimum catalogue (v1.0) (2025). \doi{10.5281/ZENODO.15878141}, \url{https://zenodo.org/doi/10.5281/zenodo.15878141}

\bibitem{onahce}
Onah, C., Firt, R., Michielsen, K.: Empirical quantum advantage in constrained optimization from encoded unitary designs. arXiv preprint arXiv:2511.14296  (2025)

\bibitem{onah2025data}
Onah, C., Guin, A., Montanez~Barrera, A., Othmer, C., Michielsen, K.: Dataset for quantum and classical approaches to the optimisation of highway platooning: the two-vehicle matching problem (2025). \doi{10.5281/ZENODO.17768136}, \url{https://zenodo.org/doi/10.5281/zenodo.17768136}

\bibitem{onah2025}
Onah, C., Misciasci, N., Othmer, C., Michielsen, K.: Quest: Quantum-enhanced shared transportation. In: 2025 IEEE International Conference on Quantum Computing and Engineering (QCE). vol.~1, pp. 2149--2160. IEEE (2025)

\bibitem{onah2025empiricaldata}
Onah, C., Roman, F., Michielsen, K.: Dataset: Empirical quantum advantage in constrained optimization from encoded unitary designs  (2025). \doi{10.5281/ZENODO.15725265}, \url{https://zenodo.org/doi/10.5281/zenodo.15725265}

\bibitem{gurobi2024}
Optimization, G., Gurobi, L.: Optimizer reference manual (2024), https://www.gurobi.com

\bibitem{ronnow2014speedup}
R{\o}nnow, T.~\, F., Wang, Z., Job, J., et~al.: Defining and detecting quantum speedup. Science  \textbf{345}(6195),  420--424 (2014)

\bibitem{vert2021benchmarking}
Vert, D., Sirdey, R., Louise, S.: Benchmarking quantum annealing against ``hard'' instances of the bipartite matching problem. SN Computer Science  \textbf{2}, ~133 (2021)

\bibitem{zhou2020quantum}
Zhou, L., Wang, S.T., Choi, S., Pichler, H., Lukin, M.~\, D.: Quantum approximate optimization algorithm: Performance, mechanism, and implementation on near-term devices. Physical Review X  \textbf{10}(2),  021067 (2020)

\end{thebibliography}

\include{supp}


\appendix

\section{Supplementary Material.}
\label{app:app}

We provide the benchmarking tables supporting the solver-zoo results in the main text. For each instance size $n\in\{3,\dots,12\}$ and each solver, the supplement reports the raw and derived metrics used in our comparisons (best/mean energy, gaps to the exact baseline $E_\star$,
success probability, STS/TTS, feasibility counts, and wall time), as well as energy-savings tables
that map matchings to $F_{\mathrm{ref}}$, $F_{\mathrm{ref}}^{\mathrm{vel}}$, $F_1$, $F_1^{\mathrm{mod}}$ and the
corresponding $\eta$-metrics.  These tables serve as a reproducible audit trail and allow readers to
reconstruct every plotted summary statistic.

\subsection{Solver Comparison Tables}
\label{app:tables}

Here we report the polynomial-time Hungarian assignment baseline on the same $n\times n$ surfer--breaker matching instances as the heuristic solvers. As an
exact algorithm, the Hungarian solver always returns a feasible match
with $p_{\text{feas}} = p_{\text{succ}} = 1$, vanishing optimality gaps,
and zero energy variance.  The wall-clock time remains essentially
negligible across $n=3,\dots,12$ compared to MIQP and all heuristic
quantum and classical samplers. For the heuristics, we report the total number of reads ($N_{\mathrm{reads}}$), number of unique samples ($N_{\mathrm{unique}}$), feasible counts in reads/unique ($n_{\mathrm{feas}}^{\mathrm{reads}}$, $n_{\mathrm{feas}}^{\mathrm{unique}}$), exact and tolerance-based success counts ($n_{\mathrm{succ,exact}}$, $n_{\mathrm{succ,tol}}$), feasibility and success probabilities ($p_{\mathrm{feas}}$, $p_{\mathrm{succ,exact}}$, $p_{\mathrm{succ,tol}}$), samples-to-solution (STS), time-to-solution (TTS), internal evaluation proxy ($N_{\mathrm{eval}}$), wall time, and energy statistics relative to the ground state $E_\star$.

\begin{table}[H]
\caption{Hungarian and Gurobi MIQP}
\label{tab:hungarian_and_miqp}
\centering
\begin{sideways}
\begin{adjustbox}{max width=\textheight, max totalheight=\textwidth}
\begin{tabular}{cccccccccccccccc}

\multicolumn{14}{c}{\textbf{Table \textbf{(a)}: Hungarian assignment baseline metrics for the WaaS instances with $n=3,\dots,12$.}} \\

\multicolumn{1}{l}{}                &            &                    &                            &                            &                              &                            &              &                      &                            &                    &                    &                       &                             &                            &                    \\ 
\cline{1-14} 
\multicolumn{1}{l}{\textbf{SOLVER}} & \textbf{n} & \textbf{$E_\star$} & \textbf{$E_{\text{best}}$} & \textbf{$E_{\text{mean}}$} & \textbf{$p_{\text{succ}}$} & \textbf{STS} & \textbf{TTS {[}s{]}}  & \textbf{gap\_best} & \textbf{gap\_mean} & \textbf{$\sigma_E^2$} & \textbf{wall\_time {[}s{]}} & \textbf{-}                  & \textbf{-}          \\ \cline{1-14} 
\multirow{10}{*}{Hungarian} & 3  & 289224.42 & 289224.42 & 289224.42 & 0 & 0 & 0 & 1 & 1 & 0 & 0 &  &  \\
          & 4  & 236096.63 & 236096.63 & 236096.63 & 0 & 0 & 0 & 1 & 1 & 0 & 0 &  &  \\
          & 5  & 415103.38 & 415103.38 & 415103.38 & 0 & 0 & 0 & 1 & 1 & 0 & 0 &  &  \\
          & 6  & 486510    & 486510    & 486510    & 0 & 0 & 0 & 1 & 1 & 0 & 0 &  &  \\
          & 7  & 511283.17 & 511283.17 & 511283.17 & 0 & 0 & 0 & 1 & 1 & 0 & 0 &  &  \\
          & 8  & 532659.67 & 532659.67 & 532659.67 & 0 & 0 & 0 & 1 & 1 & 0 & 0 &  &  \\
          & 9  & 485653.17 & 485653.17 & 485653.17 & 0 & 0 & 0 & 1 & 1 & 0 & 0 &  &  \\
          & 10 & 578870.88 & 578870.88 & 578870.88 & 0 & 0 & 0 & 1 & 1 & 0 & 0 &  &  \\
          & 11 & 722931.75 & 722931.75 & 722931.75 & 0 & 0 & 0 & 1 & 1 & 0 & 0 &  &  \\
          & 12 & 761017.5  & 761017.5  & 761017.5  & 0 & 0 & 0 & 1 & 1 & 0 & 0 &  & \\ \cline{1-14} 
\hline
\\[1cm]  
\multicolumn{14}{c}{\textbf{Table \textbf{(b)}: Gurobi MIQP baseline metrics for the surfer--breaker matching instances with $n=3,\dots,12$.}} \\
\multicolumn{1}{l}{}                &            &                    &                            &                            &                              &                            &              &                      &                            &                    &                    &                       &                             &                            &                    \\ \cline{1-14} 
\multicolumn{1}{l}{\textbf{SOLVER}} & \textbf{n} & \textbf{$E_\star$} & \textbf{$E_{\text{best}}$} & \textbf{$E_{\text{mean}}$} & \textbf{$p_{\text{succ}}$} & \textbf{STS} & \textbf{TTS {[}s{]}}  & \textbf{gap\_best} & \textbf{gap\_mean} & \textbf{$\sigma_E^2$} & \textbf{wall\_time  {[}s{]}} & \textbf{$E_{\mathrm{LB}}$} & \textbf{miqp\_gap} \\ \cline{1-14} 
\multirow{10}{*}{Gurobi-MIQP} & 3  & -664153   & -664153   & -664153   & 0 & 0 & 0 & 1 & 1 & 0.0061329 & 0.00613291 & -664153   & 0        \\
            & 4  & -705978   & -705978   & -705978   & 0 & 0 & 0 & 1 & 1 & 0.0038813 & 0.00388131 & -705978   & 0        \\
            & 5  & -3.84E+06 & -3.84E+06 & -3.84E+06 & 0 & 0 & 0 & 1 & 1 & 0.0043231 & 0.00432311 & -3.84E+06 & 0        \\
            & 6  & -3.38E+06 & -3.38E+06 & -3.38E+06 & 0 & 0 & 0 & 1 & 1 & 0.00586   & 0.00586001 & -3.38E+06 & 0        \\
            & 7  & -5.91E+06 & -5.91E+06 & -5.91E+06 & 0 & 0 & 0 & 1 & 1 & 0.008417  & 0.00841702 & -5.91E+06 & 0        \\
            & 8  & -7.95E+06 & -7.95E+06 & -7.95E+06 & 0 & 0 & 0 & 1 & 1 & 0.0112449 & 0.0112449  & -7.95E+06 & 0        \\
            & 9  & -1.01E+07 & -1.01E+07 & -1.01E+07 & 0 & 0 & 0 & 1 & 1 & 0.434102  & 0.434102   & -1.01E+07 & 0        \\
            & 10 & -1.28E+07 & -1.28E+07 & -1.28E+07 & 0 & 0 & 0 & 1 & 1 & 0.755872  & 0.755872   & -1.28E+07 & 0        \\
            & 11 & -2.19E+07 & -2.19E+07 & -2.19E+07 & 0 & 0 & 0 & 1 & 1 & 1         & 1          & -2.19E+07 & 0        \\
            & 12 & -2.61E+07 & -2.61E+07 & -2.61E+07 & 0 & 0 & 0 & 1 & 1 & 3         & 3          & -2.61E+07 & 3.83E-05 \\ 
\cline{1-14} 
\end{tabular}
\end{adjustbox}
\end{sideways}

\end{table}





\begin{table}[H]
\caption{Simulated Annealing and Tabu Search}
\label{tab:sa_and_tabu}
\centering
\begin{sideways}
\begin{adjustbox}{max width=\textheight, max totalheight=\textwidth}

\begin{tabular}{cccccccccccccccccccccccc}

\multicolumn{22}{c}{\textbf{Table \textbf{(a)}: Simulated annealing (SA) benchmarking summary on the surfer-breaker QUBO instances with $n=3,\dots,12$.}} \\

\multicolumn{1}{l}{}                &            &                    &                            &                            &                              &                            &              &                      &                            &                    &                    &                       &                             &                            &                    \\ \cline{1-22} 

\textbf{SOLVER}        & \textbf{n} & \textbf{$E_\star$} & \textbf{$E_{\text{best}}$} & \textbf{$E_{\text{mean}}$} & \textbf{p\_succ,total} & \textbf{STS\_tot} & \textbf{TTS {[}s{]}}  & \textbf{gap\_best} & \textbf{gap\_mean} & \textbf{$\sigma_E^2$} & \textbf{wall\_time {[}s{]}} & \textbf{$N_{\mathrm{reads}}$} & \textbf{$N_{\mathrm{unique}}$} & \textbf{$n_{\mathrm{feas}}^{\mathrm{reads}}$} & \textbf{$n_{\mathrm{feas}}^{\mathrm{reads}}$} & \textbf{$n_{\mathrm{succ,exact}}^{\mathrm{reads}}$} & \textbf{$n_{\mathrm{succ,exact}}^{\mathrm{unique}}$} & \textbf{$n_{\mathrm{succ,tol}}^{\mathrm{reads}}$} & \textbf{$n_{\mathrm{succ,tol}}^{\mathrm{unique}}$} & \textbf{$p_{\mathrm{succ,exact}}$} \\
 \\ \cline{1-22} 
\multirow{10}{*}{SA} & 3  & -664153   & -664153   & -642098   & 2.51E-06  & 3.32083 & 2.65E+08 & 0.275556 & 3.62903  & 0.260133 & 0.260133 & 1350  & 6     & 1350  & 6     & 372 & 1 & 372 & 1  & 1        & 0.275556 \\
   & 4  & -705978   & -696790   & -678267   & 1.30152   & 3.92522 & 2.47E+08 & 0        & $\infty$ & $\infty$ & 1.11575  & 3200  & 51    & 2013  & 24    & 0   & 0 & 0   & 0  & 0.629062 & 0        \\
   & 5  & -3.84E+06 & -3.84E+06 & -3.73E+06 & 8.69E-06  & 2.78552 & 4.60E+09 & 0.0424   & 23.5849  & 3.54369  & 3.54369  & 6250  & 120   & 6250  & 120   & 140 & 1 & 265 & 2  & 1        & 0.0224   \\
   & 6  & -3.38E+06 & -3.38E+06 & -3.29E+06 & -8.14E-06 & 2.77748 & 4.64E+09 & 0.013611 & 73.4694  & 8.47563  & 8.47563  & 10800 & 721   & 10799 & 720   & 101 & 2 & 147 & 3  & 0.999907 & 0.009352 \\
   & 7  & -5.91E+06 & -5.91E+06 & -5.75E+06 & 3.25E-06  & 2.65654 & 4.80E+09 & 0.001516 & 659.615  & 3.30959  & 18.6996  & 17150 & 4600  & 17150 & 4600  & 26  & 2 & 26  & 2  & 1        & 0.001516 \\
   & 8  & -7.95E+06 & -7.95E+06 & -7.73E+06 & -7.33E-07 & 2.84316 & 8.72E+09 & 0.000664 & 1505.88  & 10.1422  & 37.4525  & 25600 & 17308 & 25600 & 17308 & 3   & 1 & 17  & 5  & 1        & 0.000117 \\
   & 9  & -1.01E+07 & -1.01E+07 & -9.88E+06 & -2.33E-05 & 2.49217 & 8.71E+09 & 0.000466 & 2144.12  & 18.6282  & 68.7822  & 36450 & 34158 & 36450 & 34158 & 4   & 4 & 17  & 13 & 1        & 0.00011  \\
   & 10 & -1.28E+07 & -1.28E+07 & -1.25E+07 & -2.30E-05 & 2.97453 & 1.53E+10 & 4.00E-05 & 25000    & 277.34   & 120.45   & 50000 & 49484 & 50000 & 49484 & 2   & 2 & 2   & 2  & 1        & 4.00E-05 \\
   & 11 & -2.20E+07 & -2.19E+07 & -2.14E+07 & 0.194748  & 2.51915 & 2.06E+10 & 0        & $\infty$ & $\infty$ & 198.022  & 66550 & 66471 & 66550 & 66471 & 0   & 0 & 0   & 0  & 1        & 0        \\
   & 12 & -2.61E+07 & -2.61E+07 & -2.56E+07 & 0.0759048 & 2.11518 & 2.54E+10 & 1.16E-05 & 86400    & 1417.56  & 307.821  & 86400 & 86386 & 86400 & 86386 & 0   & 0 & 1   & 1  & 1        & 0     \\\cline{1-22} 
\hline
\\[1cm]  

\multicolumn{22}{c}{\textbf{Table \textbf{(b)}: Tabu search baseline metrics for the surfer--breaker
  matching instances with $n=3,\dots,12$.}} \\

\multicolumn{1}{l}{}                &            &                    &                            &                            &                              &                            &              &                      &                            &                    &                    &                       &                             &                            &                    \\ \cline{1-22} 
\textbf{SOLVER}        & \textbf{n} & \textbf{$E_\star$} & \textbf{$E_{\text{best}}$} & \textbf{$E_{\text{mean}}$} & \textbf{p\_succ,total} & \textbf{STS\_tot} & \textbf{TTS {[}s{]}}  & \textbf{gap\_best} & \textbf{gap\_mean} & \textbf{$\sigma_E^2$} & \textbf{wall\_time {[}s{]}} & \textbf{$N_{\mathrm{reads}}$} & \textbf{$N_{\mathrm{unique}}$} & \textbf{$n_{\mathrm{feas}}^{\mathrm{reads}}$} & \textbf{$n_{\mathrm{feas}}^{\mathrm{reads}}$} & \textbf{$n_{\mathrm{succ,exact}}^{\mathrm{reads}}$} & \textbf{$n_{\mathrm{succ,exact}}^{\mathrm{unique}}$} & \textbf{$n_{\mathrm{succ,tol}}^{\mathrm{reads}}$} & \textbf{$n_{\mathrm{succ,tol}}^{\mathrm{unique}}$} & \textbf{$p_{\mathrm{succ,exact}}$} \\
 \\ \cline{1-22} 
\multirow{10}{*}{TABU} & 3  & -664153   & -664153   & -664153   & 2.51E-06  & 2.51E-06  & 1.36E-20 & 1        & 1        & 28.3509  & 28.3509 & 1350  & 1    & 1350  & 1    & 1350  & 1 & 1350  & 1   & 1 & 1        \\
     & 4  & -705978   & —         & —         & —         & —         & —        & 0        & $\infty$ & $\infty$ & 67.2024 & 3200  & 2    & 0     & 0    & 0     & 0 & 0     & 0   & 0 & 0        \\
     & 5  & -3.84E+06 & -3.84E+06 & -3.84E+06 & 8.69E-06  & 8.69E-06  & 8.67E-19 & 1        & 1        & 131.26   & 131.26  & 6250  & 1    & 6250  & 1    & 6250  & 1 & 6250  & 1   & 1 & 1        \\
     & 6  & -3.38E+06 & -3.38E+06 & -3.38E+06 & -8.14E-06 & -8.14E-06 & 2.17E-19 & 1        & 1        & 226.811  & 226.811 & 10800 & 2    & 10800 & 2    & 10800 & 2 & 10800 & 2   & 1 & 1        \\
     & 7  & -5.91E+06 & -5.91E+06 & -5.90E+06 & 3.25E-06  & 0.051874  & 2.33E+07 & 0.677668 & 1.47565  & 360.176  & 360.176 & 17150 & 30   & 17150 & 30   & 11622 & 2 & 11622 & 2   & 1 & 0.677668 \\
     & 8  & -7.95E+06 & -7.95E+06 & -7.95E+06 & -7.33E-07 & 0.0294433 & 5.78E+06 & 0.977539 & 1.02298  & 537.657  & 537.657 & 25600 & 19   & 25600 & 19   & 11972 & 1 & 25025 & 6   & 1 & 0.467656 \\
     & 9  & -1.01E+07 & -1.01E+07 & -1.01E+07 & -2.33E-05 & 0.337737  & 3.07E+08 & 0.122222 & 8.18182  & 765.553  & 765.553 & 36450 & 1850 & 36450 & 1850 & 831   & 4 & 4455  & 18  & 1 & 0.022798 \\
     & 10 & -1.28E+07 & -1.28E+07 & -1.28E+07 & -2.30E-05 & 0.173814  & 1.55E+08 & 0.23528  & 4.25026  & 1050.19  & 1050.19 & 50000 & 440  & 50000 & 440  & 2405  & 2 & 11764 & 11  & 1 & 0.0481   \\
     & 11 & -2.20E+07 & -2.19E+07 & -2.19E+07 & 4.94E-07  & 0.175365  & 3.08E+08 & 0.177295 & 5.64031  & 1397.87  & 1397.87 & 66550 & 648  & 66550 & 648  & 1111  & 1 & 11799 & 10  & 1 & 0.016694 \\
     & 12 & -2.61E+07 & -2.61E+07 & -2.61E+07 & -9.57E-08 & 0.129737  & 1.46E+08 & 0.27015  & 3.70164  & 1814.92  & 1814.92 & 86400 & 4320 & 86400 & 4320 & 15    & 2 & 23341 & 315 & 1 & 0.000174\\   \cline{1-22} 
\end{tabular}
\end{adjustbox}
\end{sideways}
\end{table}

\begin{table}[H]
\caption{Leap hybrid Solver, QPU-QA and CE-QAOA}
\label{tab:qpu_leap_ceqaoa}
\centering
\begin{sideways}
\begin{adjustbox}{max width=\textheight, max totalheight=\textwidth}

\begin{tabular}{ccccccccccccccccc}

\multicolumn{15}{c}{\textbf{Table \textbf{(a)}: Leap hybrid solver baseline metrics for the surfer--breaker matching instances with $n=3,\dots,12$.}} \\

\multicolumn{1}{l}{}                &            &                    &                            &                            &                              &                            &              &                      &                            &                    &                    &                       &                             &                            &                    \\ \cline{1-15} 
\textbf{SOLVER}               & \textbf{n} & \textbf{$E_\star$} & \textbf{$E_{\text{best}}$} & \textbf{$E_{\text{mean}}$} & \textbf{$p_{\text{succ}}$} & \textbf{STS} & \textbf{TTS {[}s{]}}  & \textbf{gap\_best} & \textbf{gap\_mean} & \textbf{$\sigma_E^2$} & \textbf{wall\_time {[}s{]}} & \textbf{}                       & \textbf{}                      & \textbf{}            \\
\cline{1-15} 
\multirow{10}{*}{LEAP} & 3  & -664153.2 & -6.64E+05 & -6.64E+05 & 0.000003  & 0.000003  & 0 & 1 & 1 & 2.544257 & 2.544257 &  &  &  \\
     & 4  & -705978.4 & -7.06E+05 & -7.06E+05 & 0.000002  & 0.000002  & 0 & 1 & 1 & 2.40487  & 2.40487  &  &  &  \\
     & 5  & -3835914  & -3.84E+06 & -3.84E+06 & 0.000009  & 0.000009  & 0 & 1 & 1 & 2.537564 & 2.537564 &  &  &  \\
     & 6  & -3379371  & -3.38E+06 & -3.38E+06 & -0.000008 & -0.000008 & 0 & 1 & 1 & 2.446772 & 2.446772 &  &  &  \\
     & 7  & -5905685  & -5.91E+06 & -5.91E+06 & 0.000003  & 0.000003  & 0 & 1 & 1 & 2.405447 & 2.405447 &  &  &  \\
     & 8  & -7953421  & -7.95E+06 & -7.95E+06 & 0.075439  & 0.075439  & 0 & 1 & 1 & 2.519187 & 2.519187 &  &  &  \\
     & 9  & -10130600 & -1.01E+07 & -1.01E+07 & 0.273869  & 0.273869  & 0 & 1 & 1 & 2.521728 & 2.521728 &  &  &  \\
     & 10 & -12847740 & -1.28E+07 & -1.28E+07 & 0.378651  & 0.378651  & 0 & 1 & 1 & 2.548852 & 2.548852 &  &  &  \\
     & 11 & -21949950 & -2.17E+07 & -2.17E+07 & 1.02324   & 1.02324   & 0 & 1 & 1 & 2.574582 & 2.574582 &  &  &  \\
     & 12 & -26125560 & -2.60E+07 & -2.60E+07 & 0.519158  & 0.519158  & 0 & 1 & 1 & 2.523256 & 2.523256 &  &  & \\\cline{1-15} 
\hline
\\[0.4cm]  

\multicolumn{15}{c}{\textbf{Table \textbf{(b)}: Quantum annealing (QPU-QA) baseline metrics for the surfer--breaker matching instances with $n=3,\dots,12$.}} \\

\multicolumn{1}{l}{}                &            &                    &                            &                            &                              &                            &              &                      &                            &                    &                    &                       &                             &                            &                    \\ \cline{1-15} 

\textbf{SOLVER}               & \textbf{n} & \textbf{$E_\star$} & \textbf{$E_{\text{best}}$} & \textbf{$E_{\text{mean}}$} & \textbf{$p_{\text{succ}}$} & \textbf{STS} & \textbf{TTS {[}s{]}}  & \textbf{gap\_best} & \textbf{gap\_mean} & \textbf{$\sigma_E^2$} & \textbf{wall\_time {[}s{]}} & \textbf{}                       & \textbf{}                      & \textbf{}            \\
\cline{1-15} 
\multirow{10}{*}{QPU-QA} & 3  & -664153.2 & -664153.2 & -4.76E+07 & -2.27E-07 & 0.718004 & 1.66E+11 & 0.013333 & 75 & 0.06907 & 0.092249 &  &  &  \\
       & 4  & –         & –         & –         & –         & –        & –        & –        & –  & –       & –        &  &  &  \\
       & 5  & –         & –         & –         & –         & –        & –        & –        & –  & –       & –        &  &  &  \\
       & 6  & –         & –         & –         & –         & –        & –        & –        & –  & –       & –        &  &  &  \\
       & 7  & –         & –         & –         & –         & –        & –        & –        & –  & –       & –        &  &  &  \\
       & 8  & –         & –         & –         & –         & –        & –        & –        & –  & –       & –        &  &  &  \\
       & 9  & –         & –         & –         & –         & –        & –        & –        & –  & –       & –        &  &  &  \\
       & 10 & –         & –         & –         & –         & –        & –        & –        & –  & –       & –        &  &  &  \\
       & 11 & –         & –         & –         & –         & –        & –        & –        & –  & –       & –        &  &  &  \\
       & 12 & –         & –         & –         & –         & –        & –        & –        & –  & –       & –        &  &  &  \\\cline{1-15} 
\hline
\\[0.4cm]  

\multicolumn{15}{c}{\textbf{Table \textbf{(c)}: CE-QAOA $p=1$ baseline metrics for the surfer--breaker
  matching instances with $n=3,\dots,12$.}} \\

\multicolumn{1}{l}{}                &            &                    &                            &                            &                              &                            &              &                      &                            &                    &                    &                       &                             &                            &                    \\ \cline{1-15} 
\textbf{SOLVER}               & \textbf{n} & \textbf{$E_\star$} & \textbf{$E_{\text{best}}$} & \textbf{$E_{\text{mean}}$} & \textbf{$p_{\text{succ}}$} & \textbf{STS} & \textbf{TTS {[}s{]}}  & \textbf{gap\_best} & \textbf{gap\_mean} & \textbf{$\sigma_E^2$} & \textbf{wall\_time {[}s{]}} & \textbf{$\gamma_{\text{best}}$} & \textbf{$\beta_{\text{best}}$} & \textbf{grid\_evals} \\
\cline{1-15} 
\multirow{10}{*}{CE-QAOA $p=1$}
& 3  & -664153.1833  & -664153.1833  & -637137.1885  & 0.000493827 & 2025               & 0.8439542405000111 & 0                    & 3.8949822024248126 & 97017351.27624512 & 1.6879084810000222 & 1.215 & 0.65  & 1 \\
& 4  & -696789.9833  & -696789.9833  & -662714.1625  & 0.006171875 & 162.02531645569618 & 0.018429363        & 0                    & 4.255877841421934  & 442787449.3477173 & 1.455919659000017  & 2.375 & 0.779 & 1 \\
& 5  & -3835913.667  & -3835913.667  & -3695132.661  & 0.000512    & 1953.125           & 0.3637791215625015 & 0                    & 3.7033476397929923 & 5474372379.517578 & 5.820465945000024  & 2.36  & 2.327 & 1 \\
& 6  & -3379371.275  & -3379371.275  & -3237746.234  & 0           & $\infty$           & $\infty$           & 0.037186404345927945 & 3.7659176601021156 & 4440013620.529297 & 36.28281790699998  & 2.068 & 0.497 & 1 \\
& 7  & -5905684.808  & -5876006.725  & -5710483.383  & 0           & $\infty$           & $\infty$           & 0.2539925594882839   & 3.731117242778161  & 5301543150.589844 & 92.18202271        & 2.795 & 0.543 & 1 \\
& 8  & -7953421.058  & -7936565.392  & -7667142.702  & 0           & $\infty$           & $\infty$           & 0.21192976636104913  & 3.5994366993384954 & 9398836487.789062 & 100.82788856100001 & 1.626 & 2.51  & 1 \\
& 9  & -10130602.36  & -10056341.69  & -9837179.529  & 0           & $\infty$           & $\infty$           & 0.7330330817456795   & 2.8964006213858173 & 8364942572.546875 & 430.85198196500005 & 0.16  & 2.295 & 1 \\
& 10 & -12847742.96  & -12637655.96  & -12357610.94  & 0           & $\infty$           & $\infty$           & 1.6352055040426057   & 3.814927053249603  & 19198801547.03125 & 401.4604331        & 1.417 & 0.853 & 1 \\
& 11 & -21949949.89  & -21773245.06  & -21336791.99  & 0           & $\infty$           & $\infty$           & 0.805035247          & 2.7934364522101824 & 21793378311       & 645.0540156        & 2.35  & 0.804 & 1 \\
& 12 & -26125560.03  & -25828688.19  & -25504276.7   & 0           & $\infty$           & $\infty$           & 1.1363271564273298   & 2.3780670286065773 & 28172280925       & 984.7312200189999  & 1.829 & 0.822 & 1 \\\cline{1-15}
\end{tabular}
\end{adjustbox}
\end{sideways}
\end{table}

\subsection{Energy Savings}
\label{app:energy_savings}

Energy benchmarks for the various solvers
across problem sizes $n=3,\dots,12$ are given in Tables \ref{tab:hungarian-four-costs}, \ref{tab:energy_qpu_leap_qaoa} and  \ref{tab:energy_miqp_sa_tabu}. For each instance size $n$ we report the
reference solo fuel cost at the surfers' preferred speeds $F_{\text{ref}}$,
the solo cost at the breakers' speeds $F_{\text{ref,vel}}$, the paired
aerodynamic cost $F_1$, the post-hoc corrected cost $F_{1,\text{mod}}$, the
relative speed efficiency $\eta_{\text{speed}}$, and the relative fuel
efficiencies $\eta_{F_1}$ and $\eta_{F_{1,\text{mod}}}$ (all in \%). The relative speed efficiency $\eta_{\text{speed}}$ measures the average deviation in speed between the paired solution and the solo reference (in percent), whereas $\eta_{F_1}$ and $\eta_{F_{1,\text{mod}}}$ measure the relative fuel savings of the paired solution with respect to $F_{\text{ref}}$. 

\begin{table}[htbp]
  \centering
  \caption{Hungarian baseline evaluation of the four aerodynamic cost functions
  for single-segment surfer--breaker matching instances with $n=3,\dots,12$ vehicles.
  }
  \label{tab:hungarian-four-costs}
  \begin{tabular}{lrrrrrrrr}
    \toprule
    \textbf{solver} & \textbf{n} & \textbf{$F_\text{ref}$} & \textbf{$F_\text{ref}^\text{vel}$} & \textbf{$F_1$} &
    \textbf{$F_1^{\text{mod}}$} & \textbf{$\eta_\text{speed}^\%$} & \textbf{$\eta_{F_1}^\%$} & \textbf{$\eta_{F_1^{\text{mod}}}^\%$} \\
    \midrule
    \multirow{10}{*}{Hungarian} & 3 & 118196 & 99965 & 85224.4 & 118196 & 15.4244 & 27.8957 & 0 \\
      & 4 & 118848 & 77313 & 59872.4 & 118848 & 34.948 & 49.6227 & 0 \\
      & 5 & 177451 & 191125 & 161813 & 177451 & -7.70579 & 8.81257 & 0 \\
      & 6 & 191983 & 128938 & 100139 & 191983 & 32.8388 & 47.8396 & 0 \\
      & 7 & 250936 & 188828 & 150047 & 250936 & 24.7505 & 40.2049 & 0 \\
      & 8 & 319735 & 275579 & 227127 & 319735 & 13.8102 & 28.9639 & 0 \\
      & 9 & 386078 & 340914 & 284404 & 386078 & 11.6982 & 26.3351 & 0 \\
      & 10 & 408253 & 375535 & 310364 & 403168 & 8.01415 & 23.9776 & 1.24563 \\ 
      & 11 & 460393 & 435895 & 361098 & 447370 & 5.32111 & 21.5674 & 2.82874 \\
      & 12 & 483876 & 373206 & 301902 & 468732 & 22.8716 & 37.6075 & 3.12976 \\
    \bottomrule
  \end{tabular}
\end{table}


\begin{table}[htbp]
\caption{Energy Saving for MIQP, SA and Tabu search}
\label{tab:energy_miqp_sa_tabu}
\centering
\begin{adjustbox}{max width=\textheight, max height=\textwidth}
\begin{tabularx}{\textwidth}{cccccccccc}
\\\cline{1-10}
\multicolumn{10}{p{\textwidth}}{\centering 
\textbf{Table (a): Fuel / energy benchmarks for the Gurobi-based MIQP }\\ 
\textbf{assignment solver across problem sizes $n=3,\dots,12$.}
}\\
\\\cline{1-10}
\textbf{SOLVER} & \textbf{n} & \textbf{$F_\text{ref}$} & \textbf{$F_\text{ref}^\text{vel}$} & \textbf{$F_1$} &  \textbf{$F_1^{\text{mod}}$} & \textbf{$\eta_\text{speed}^\%$} & \textbf{$\eta_{F_1}^\%$} & \textbf{$\eta_{F_1^{\text{mod}}}^\%$} & \textbf{wall\_time {[}s{]}} \\

\multirow{10}{*}{MIQP} & 3  & 118196 &  99965 &  85224.4 & 118196 & 15.4244 & 27.8957 & 0       & 0.00320431 \\
  & 4  & 118848 &  77313 &  59872.4 & 118848 & 34.948  & 49.6227 & 0       & 0.00319371 \\
  & 5  & 177451 & 191125 & 161813   & 177451 & -7.70579&  8.81257& 0       & 0.00418791 \\
  & 6  & 191983 & 128938 & 100139   & 191983 & 32.8388 & 47.8396 & 0       & 0.00409631 \\
  & 7  & 250936 & 188828 & 150047   & 250936 & 24.7505 & 40.2049 & 0       & 0.00498731 \\
  & 8  & 319735 & 275579 & 227127   & 319735 & 13.8102 & 28.9639 & 0       & 0.00622701 \\
  & 9  & 386078 & 340914 & 284404   & 386078 & 11.6982 & 26.3351 & 0       & 0.00897401 \\
  & 10 & 408253 & 375535 & 310364   & 403168 &  8.01415& 23.9776 & 1.24563 & 0.00887621 \\
  & 11 & 460393 & 435895 & 361098   & 447370 &  5.32111& 21.5674 & 2.82874 & 0.0109073  \\
  & 12 & 483876 & 373206 & 301902   & 468732 & 22.8716 & 37.6075 & 3.12976 & 0.0131957  \\

\\\cline{1-10} 
\\\cline{1-10} 
\multicolumn{10}{p{\textwidth}}{\centering 
\textbf{Table (b): Fuel / energy benchmarks for the Simulated Annealing}\\ 
\textbf{solver across problem sizes $n=3,\dots,12$.}
}\\
\\\cline{1-10}
\textbf{SOLVER} & \textbf{n} & \textbf{$F_\text{ref}$} & \textbf{$F_\text{ref}^\text{vel}$} & \textbf{$F_1$} &  \textbf{$F_1^{\text{mod}}$} & \textbf{$\eta_\text{speed}^\%$} & \textbf{$\eta_{F_1}^\%$} & \textbf{$\eta_{F_1^{\text{mod}}}^\%$} & \textbf{wall\_time {[}s{]}} \\ 

\multirow{10}{*}{SA} & 3  & 118196 & 90259  & 73225.17  & 118196   & 23.63616   & 38.04768  & 0        & 0.291408 \\
  & 4  & ---    & ---    & ---      & ---    & ---     & ---     & ---    & --- \\
  & 5  & 177451 & 237021 & 208184.5  & 177451   & -33.56983  & -17.31941 & 0        & 3.721341 \\
  & 6  & 191983 & 153819 & 125088.5  & 191983   & 19.87884   & 34.84395  & 0        & 9.310699 \\
  & 7  & 250936 & 188828 & 150047.5  & 250936   & 24.75053   & 40.20489  & 0        & 20.60184 \\
  & 8  & 319735 & 308982 & 261156.2  & 319735   & 3.363098   & 18.32104  & 0        & 41.10357 \\
  & 9  & 386078 & 358159 & 302137    & 386078   & 7.23144    & 21.74197  & 0        & 75.49417 \\
  & 10 & 408253 & 403493 & 337932.7  & 408253   & 1.165944   & 17.22469  & 0        & 130.9877 \\
  & 11 & 460393 & 449256 & 373756.5  & 452070.2 & 2.41902    & 18.81795  & 1.807767 & 216.8527 \\
  & 12 & 483876 & 415056 & 342883.5  & 483876   & 14.22265   & 29.13815  & 0        & 339.4152 \\
\\\cline{1-10} 
\\\cline{1-10}
\multicolumn{10}{p{\textwidth}}{\centering 
\textbf{Table (c): Fuel / energy benchmarks for the Tabu search}\\ 
\textbf{across problem sizes $n=3,\dots,12$.}
}\\
\\\cline{1-10}
\textbf{SOLVER} & \textbf{n} & \textbf{$F_\text{ref}$} & \textbf{$F_\text{ref}^\text{vel}$} & \textbf{$F_1$} &  \textbf{$F_1^{\text{mod}}$} & \textbf{$\eta_\text{speed}^\%$} & \textbf{$\eta_{F_1}^\%$} & \textbf{$\eta_{F_1^{\text{mod}}}^\%$} & \textbf{wall\_time {[}s{]}} \\ 

\multirow{10}{*}{TABU} & 3  & 118196 & 90259  & 73225.17  & 118196   & 23.63616   & 38.04768  & 0        & 28.35091 \\
  & 4  & ---    & ---    & ---      & ---    & ---     & ---     & ---    & --- \\
  & 5  & 177451 & 237021 & 208184.5  & 177451   & -33.56983  & -17.31941 & 0        & 131.2603 \\
  & 6  & 191983 & 153819 & 125088.5  & 191983   & 19.87884   & 34.84395  & 0        & 226.8114 \\
  & 7  & 250936 & 189328 & 149823.5  & 250936   & 24.55128   & 40.29416  & 0        & 360.1759 \\
  & 8  & 319735 & 308982 & 261156.2  & 319735   & 3.363098   & 18.32104  & 0        & 537.6573 \\
  & 9  & 386078 & 360293 & 304482.8  & 386078   & 6.678702   & 21.13437  & 0        & 765.553  \\
  & 10 & 408253 & 404429 & 338868.7  & 408253   & 0.9366741  & 16.99543  & 0        & 1050.191 \\
  & 11 & 460393 & 461378 & 386723.4  & 452070.2 & -0.2139476 & 16.00147  & 1.807767 & 1397.873 \\
  & 12 & 483876 & 419391 & 347627.9  & 483876   & 13.32676   & 28.15764  & 0        & 1814.92  \\

\\\cline{1-10} 
\end{tabularx}
\end{adjustbox}
\end{table}

\begin{table}[H]
\caption{Energy Saving for Leap hybrid solver, QPU-QA and CE-QAOA}
\label{tab:energy_qpu_leap_qaoa}
\centering
\begin{adjustbox}{max width=\textheight, max height=\textwidth}
\begin{tabularx}{\textwidth}{cccccccccc}
\\\cline{1-10} 
\multicolumn{10}{p{\textwidth}}{\centering 
\textbf{Table (a): Fuel / energy benchmarks for the Leap solver}\\ 
\textbf{across problem sizes $n=3,\dots,12$.}
}\\
\\\cline{1-10} 
\textbf{SOLVER} & \textbf{n} & \textbf{$F_\text{ref}$} & \textbf{$F_\text{ref}^\text{vel}$} & \textbf{$F_1$} &  \textbf{$F_1^{\text{mod}}$} & \textbf{$\eta_\text{speed}^\%$} & \textbf{$\eta_{F_1}^\%$} & \textbf{$\eta_{F_1^{\text{mod}}}^\%$} & \textbf{wall\_time {[}s{]}} \\ 

\multirow{10}{*}{LEAP} & 3  & 118196 &  99965 &  85224.4 & 118196 &  15.4244  & 27.8957  & 0.0000 & 0.151 \\
  & 4  & ---    & ---    & ---      & ---    & ---       & ---      & ---    & 1.201 \\
 & 5  & 177451 & 222350 & 192307.0 & 177451 & -25.3022  & -8.37208 & 0.0000 & 2.113 \\
 & 6  & 191983 & 152266 & 123548.0 & 191983 &  20.6878  & 35.6463  & 0.0000 & 48.211 \\
 & 7  & 250936 & 209134 & 170072.0 & 250936 &  16.6584  & 32.2251  & 0.0000 & 68.707 \\
 & 8  & 319735 & 291539 & 241886.0 & 319735 &   8.81855 & 24.3479  & 0.0000 & 3499.81 \\
 & 9  & 386078 & 349768 & 293105.0 & 386078 &   9.40484 & 24.0813  & 0.0000 & 113.664 \\
 & 10 & 408253 & 382559 & 316148.0 & 408253 &   6.29365 & 22.5607  & 0.0000 & 594.469 \\
 & 11 & 460393 & 478160 & 403729.0 & 460393 &  -3.85909 & 12.3078  & 0.0000 & 575.165 \\
 & 12 & 483876 & 383856 & 309568.0 & 457463 &  20.6706  & 36.0233  & 1.4586 & 1623.56 \\

\\\cline{1-10} 
\\\cline{1-10} 

\multicolumn{10}{p{\textwidth}}{\centering 
\textbf{Table (b): Fuel / energy benchmarks for the QPU solver}\\ 
\textbf{across problem sizes $n=3,\dots,12$.}
}\\
\\\cline{1-10} 
\textbf{SOLVER} & \textbf{n} & \textbf{$F_\text{ref}$} & \textbf{$F_\text{ref}^\text{vel}$} & \textbf{$F_1$} &  \textbf{$F_1^{\text{mod}}$} & \textbf{$\eta_\text{speed}^\%$} & \textbf{$\eta_{F_1}^\%$} & \textbf{$\eta_{F_1^{\text{mod}}}^\%$} & \textbf{wall\_time {[}s{]}} \\

\multirow{10}{*}{QPU-QA} & 3  & 118196 &  99965 &  85224.4 & 118196 & 15.4244 & 27.8957 & 0.0000 & 0.151 \\
  & 4  & ---    & ---    & ---      & ---    & ---     & ---     & ---    & 1.201 \\
  & 5  & ---    & ---    & ---      & ---    & ---     & ---     & ---    & 2.113 \\
  & 6  & ---    & ---    & ---      & ---    & ---     & ---     & ---    & 48.211 \\
  & 7  & ---    & ---    & ---      & ---    & ---     & ---     & ---    & 68.707 \\
  & 8  & ---    & ---    & ---      & ---    & ---     & ---     & ---    & 3499.81 \\
  & 9  & ---    & ---    & ---      & ---    & ---     & ---     & ---    & 113.664 \\
  & 10 & ---    & ---    & ---      & ---    & ---     & ---     & ---    & 594.469 \\
  & 11 & ---    & ---    & ---      & ---    & ---     & ---     & ---    & 575.165 \\
  & 12 & ---    & ---    & ---      & ---    & ---     & ---     & ---    & 1623.56 \\

\\\cline{1-10} 
\\\cline{1-10} 
\multicolumn{10}{p{\textwidth}}{\centering 
\textbf{Table (c): Fuel / energy benchmarks for the CE-QAOA solver}\\ 
\textbf{across problem sizes $n=3,\dots,12$.}
}\\
\\\cline{1-10} 
\textbf{SOLVER} & \textbf{n} & \textbf{$F_\text{ref}$} & \textbf{$F_\text{ref}^\text{vel}$} & \textbf{$F_1$} &  \textbf{$F_1^{\text{mod}}$} & \textbf{$\eta_\text{speed}^\%$} & \textbf{$\eta_{F_1}^\%$} & \textbf{$\eta_{F_1^{\text{mod}}}^\%$} & \textbf{wall\_time {[}s{]}} \\ 

\multirow{10}{*}{CE--QAOA} & 3  & 118196 & 90259  & 73225.17  & 118196.00 & 23.64  & 38.05  & 0.00 & 0.15 \\
                           & 4  & 118848 & 84589  & 67857.75  & 118848.00 & 28.83  & 42.90  & 0.00 & 1.20 \\
                           & 5  & 177451 & 237021 & 208184.46 & 177451.00 & -33.57 & -17.32 & 0.00 & 2.11 \\
                           & 6  & 191983 & 135636 & 106395.79 & 191983.00 & 29.35  & 44.58  & 0.00 & 48.21 \\
                           & 7  & 250936 & 188828 & 150047.46 & 250936.00 & 24.75  & 40.20  & 0.00 & 68.71 \\
                           & 8  & 319735 & 308982 & 261156.21 & 319735.00 & 3.36   & 18.32  & 0.00 & 3499.81 \\
                           & 9  & 386078 & 350520 & 295190.13 & 382840.50 & 9.21   & 23.54  & 0.84 & 113.66 \\
                           & 10 & 408253 & 399790 & 334886.96 & 408253.00 & 2.07   & 17.97  & 0.00 & 594.47 \\
                           & 11 & 460393 & 471774 & 396832.38 & 457155.50 & -2.47  & 13.81  & 0.70 & 575.17 \\
                           & 12 & 483876 & 406653 & 333722.33 & 478039.17 & 15.96  & 31.03  & 1.21 & 1623.56 \\

\\\cline{1-10} 
\end{tabularx}
\end{adjustbox}
\end{table}


\subsection*{Parameter Transfer in CE--QAOA}
We implemented \emph{one-shot parameter transfer} as follows:  for each instance size $n$ we run CE--QAOA at $p=1$
using a \emph{fixed} angle pair $(\gamma^\star,\beta^\star)$ imported from a generic $p=1$ QAOA baseline,
i.e., \emph{without} any per-instance variational tuning (here $\texttt{grid\_evals}=1$).  We then sample the
resulting circuit, filter to feasible outcomes, and report the best feasible bitstring $x_{\text{best}}$ by
minimum energy.  Empirically, this transfer already produces near-optimal solutions. CE-QAOA results in Table~\ref{tab:energy_qpu_leap_qaoa}
shows exact optimal hits for $n\in\{3,4,5\}$ (gap$_{\text{best}}=0$), and small approximation gaps for larger
instances despite zero re-optimization.  In particular, the best feasible energies satisfy
$\mathrm{gap}_{\text{best}}<1\%$ for $n=6,7,8,9$ (0.04\%, 0.25\%, 0.21\%, 0.73\%, respectively) and remain close
to optimal more broadly, with worst-case $\mathrm{gap}_{\text{best}}\le 1.64\%$ over $n\le 12$.  Conceptually,
the transferred angles act as a coarse but effective ``phase setting'' for the CE kernel. The encoder and block-XY mixer confine dynamics to the one-hot manifold, the imported cost-phase still induces a
useful phase gradient over low-penalty/low-cost regions, and the subsequent mixing concentrates weight on
high-quality feasible assignments.

\end{document}